\documentclass[12pt,preprint]{aastex}
\usepackage{natbib,graphics,graphicx}

\slugcomment{\date}

\shorttitle{TBD}
\shortauthors{TBD}

\begin{document}

\newcommand{\civ}{\ion{C}{4}}
\newcommand{\alphauv}{$\alpha_{\rm uv}$}
\newcommand{\aox}{$\alpha_{\rm ox}$}
\newcommand{\daox}{$\Delta \alpha_{\rm ox}$}
\newcommand{\ax}{$\alpha_{\rm x}$}
\newcommand{\HR}{{HR}}
\newcommand{\GHR}{$\Gamma_{\rm HR}$}
\newcommand{\flux}{\mbox{\,ergs~cm$^{-2}$~s$^{-1}$}}
\newcommand{\fnu}{\mbox{\,ergs~cm$^{-2}$~s$^{-1}$~Hz$^{-1}$}}
\newcommand{\lumin}{\mbox{\,ergs~s$^{-1}$}}

\title{
C\,{\sc IV} Emission and the Ultraviolet through X-ray Spectral Energy Distribution of Radio-Quiet Quasars
}

\author{
Nicholas E.\ Kruczek,\altaffilmark{1}
Gordon T.\ Richards,\altaffilmark{1,2}
S.~C. Gallagher,\altaffilmark{3}
Rajesh P.\ Deo,\altaffilmark{1,3}
Patrick B.\ Hall,\altaffilmark{4},
Paul C.\ Hewett,\altaffilmark{5}
Karen M.\ Leighly,\altaffilmark{6}
Coleman M.\ Krawczyk,\altaffilmark{1}
and
Daniel Proga\altaffilmark{7}
}

\altaffiltext{1}{Department of Physics, Drexel University, 3141 Chestnut Street, Philadelphia, PA 19104, USA}
\altaffiltext{2}{Alfred P. Sloan Research Fellow.}
\altaffiltext{3}{Department of Physics \& Astronomy, The University of Western Ontario, London, ON N6A 3K7, Canada.}
\altaffiltext{4}{Department of Physics and Astronomy, York University, Toronto, Ontario M3J 1P3, Canada.}
\altaffiltext{5}{Institute of Astronomy, Madingley Road, Cambridge CB3 0HA, UK.}
\altaffiltext{6}{Homer L. Dodge Department of Physics and Astronomy, The University of Oklahoma, 440 W. Brooks St., Norman, OK 73019, USA}
\altaffiltext{7}{Department of Physics and Astronomy, University of Nevada, Las Vegas, Box 454002, 4505 Maryland Pkwy, Las Vegas, NV 891541, USA}

\begin{abstract} 

In the restframe ultra-violet (UV), two of the parameters that best characterize the range of emission-line properties in quasar broad emission-line regions are the equivalent width and the blueshift of the \civ\ $\lambda$\,1549 line relative to the quasar rest frame.  We explore the connection between these emission-line properties and the UV through X-ray spectral energy distribution (SED) for radio-quiet (RQ) quasars.  Our sample consists of a heterogeneous compilation of 406 quasars from the Sloan Digital Sky Survey (at $z>1.54$) and Palomar-Green survey (at $z<0.4$) that have well-measured \civ\ emission-line and X-ray properties (including 164 objects with measured $\Gamma$).  We find that RQ quasars with both strong \civ\ emission and small \civ\ blueshifts can be classified as ``hard-spectrum'' sources that are (relatively) strong in the X-ray as compared to the UV.
On the other hand, RQ quasars with both weak \civ\ emission and large \civ\ blueshifts are instead ``soft-spectrum'' sources that are (relatively) weak in the X-ray as compared to the UV.  This work helps to further bridge optical/soft X-ray ``Eigenvector 1'' relationships to the UV and hard X-ray.
Based on these findings, we argue that future work should consider systematic errors in bolometric corrections (and thus accretion rates) that are derived from a single mean SED.  Detailed analysis of the \civ\ emission line may allow for SED-dependent corrections to these quantities.

\end{abstract}

\section{Introduction}

While the broad emission-line region (BELR) in active galactic nuclei (AGN) remains an enigma, in recent years there has been considerable focus on radiation line driven and magneto hydrodynamic driven winds from an accretion disk \citep[e.g.,][]{bp82,kk94,mcgv95,psk00,everett05}
as a key component.  The existence of such winds has been demonstrated for accretion disks in cataclysmic variable star systems and hydrodynamic simulations show that the phenomenon scales to the much larger quasar black hole masses \citep[e.g.,][]{psk00}; however, the detailed gas and kinematic structure of the wind remains uncertain.  This is an acute problem, which is important to our understanding of black hole growth and its influence on the evolution of the host galaxy, especially if such winds are a significant source of AGN ``feedback'' \citep[e.g.,][]{hhc+06,cop10}.

Perhaps the most important/most common wind diagnostic is the emission-line blueshift \citep{gas82}, which we have studied for a large sample of quasars \citep{rvr+02,Richards2011} and in the context of the narrow-line Seyfert 1 (NLS1) class \citep{lei01,Leighly04}.  Specifically, high-ionization emission lines, like \civ\ $\lambda$\,1549 are found to peak at wavelengths blueward of their expected laboratory wavelength when referenced to low ionization lines (which are more indicative of the quasar systemic redshift; e.g., \citealt{tf92,HW10}).  As far back as the early 1980s, this ``\ion{C}{4} blueshift'' had been suggested as the signature of an outflow \citep{gas82,Wilkes84}.  While the exact physical origin of this blueshift is not agreed upon \citep{gaskell09}, one possibility is that it is related to the strength of the radiation pressure compared with the black hole mass and is sensitive to the Eddington accretion rate \citep{sbm+07}.  
It may be that NLS1s are the low-luminosity, local analogues of large CIV blueshift quasars.  If the typically steep hard-band photon indices of NLS1s ($\Gamma>2$) are signatures of high Eddington accretion rates \citep{pdo95,bme97,Leighly99},
then the same might be the case for large CIV blueshift quasars \citep[e.g.,][]{wwz+11} since a high accretion rate for a given mass is expected to be more conducive to wind driving through radiation pressure on spectral lines \citep{psd98}.  Indeed, the common link between large \civ\ blueshift quasars, broad absorption line quasars (BALQSOs; \citealt{wmf+91}), and traditionally defined NLS1s \citep{op85} is likely a strong radiation-pressure driven wind as all three categories exhibit wind features in emission and/or absorption \citep{rvr+02,rrh+03,Leighly04} and may tend toward high $L/L_{\rm Edd}$ \citep[e.g.,][]{bg92,bor02,gbc+07}.

The premise of a radiation line driven wind \citep[e.g.,][]{ls70,cak75} is fairly simple and has applications to quasars \citep[e.g.,][]{mcgv95,psk00}.  The integrated opacity in ultra-violet (UV) bound-bound transitions accelerates gas as photons at specific wavelengths are scattered by ions in the gas.  As the gas velocity increases, the Doppler shift makes photons in new wavelength ranges able to contribute to the acceleration.  This process allows a large amount of momentum to be extracted from continuum and emission-line photons, thus driving an outflow from the accretion disk.
For this process to work, the quasar must reduce its X-ray continuum such that 
the ions which absorb in the UV are not destroyed by over-ionization.
To some extent, we know that outflows driven at least in part by radiation pressure must be occurring in BALQSOs because we see large equivalent width absorption troughs where both energy and momentum have been removed from the continuum \citep{wmf+91,akb+95}.  

In this paper, we follow-up on the work of \citet{rvr+02}, \citet{grh+05}, and \citet{Richards2011} using SDSS quasars at $z>1.54$ where the presence of \civ\ emission in the optical provides a powerful diagnostic of accretion disk physics.  In particular, \citet{Richards2011} showed that a number of different emission-line features were consistent with a two-component disk+wind model of the BELR \citep[e.g.,][]{cdm+88,Leighly04,Collin06}.  Emission-line features suggestive of a weaker than average ionizing spectrum (relative to the UV) were found in quasars with weak, highly blueshifted \civ\ emission.  On the other hand, quasars with emission-line features that indicate a strong ionizing spectrum were found to have very strong \civ\ emission lines at the systemic redshift.  
These extrema can be crudely categorized respectively as being dominated by a ``wind'' component and by a ``disk'' component of the BELR, although it seems that both components may be present in all objects.  While some of the differences seen in the BELR are likely derived from intrinsic differences in the spectral energy distribution (SED; e.g., a large {\em intrinsic} UV to X-ray flux ratio is needed to produce a strong radiation line driven wind; \citealt{mcgv95,pk04}), processing of the ionizing continuum {\em through} the wind can have a significant affect on the emission-line features that arise within the disk component \citep{Leighly04,lc07}.  Here we explore the X-ray properties of quasars (such as the optical/UV to X-ray flux ratio, \aox) over the full range of \civ\ properties to see if the evidence that the extrema in \civ\ emission-line properties traces extrema in the ionizing continuum is consistent with the X-ray properties of quasars.

While our work focuses specifically on the \civ\ emission line, our results should be viewed in the context of the broader ``eigenvector 1 (EV1)'' literature that stemmed from the ground-breaking work of \citet{bg92}.  It is beyond the scope of this paper to provide a full review of this literature, but a newcomer to the field would be well served by a review of \citet{smd00} and \citet{bf99} and references therein; Table~2 in the latter nicely summarizes a broad range of correlated emission and continuum properties of quasars.  While much of the EV1 work is founded in low-redshift quasars and investigations of optical emission lines, investigations that bridge the gap to high-redshift quasars and UV emission lines include \citet{msd+96}, \citet{wzg96}, and \citet{wlb+99}.  More recent work focused on \civ\ includes \citet{bms+04}, \citet{bl04}, \citet{bl05}, \citet{sbm+07}, \citet{msn+10}, and \citet{wwz+11}.  Lastly, \citet{Green96} and \citet{lfe+97} are among those notable for making connections between optical/UV emission lines and X-ray properties of quasars.

This paper is organized as follows. In Section~\ref{sec:data} we describe the data used in our analysis.  In Section~\ref{sec:analysis} we investigate the range of X-ray properties as a function of \civ\ emission-line parameters.  In Section~\ref{sec:discussion} we discuss the resulting SEDs.  Finally, our conclusions are presented in Section~\ref{sec:conclusions}.  Equivalent widths are given in units of rest-frame \AA, and velocities in km\,s$^{-1}$ with a sign convention such that positive velocities represent outflows in the quasar frame.  Spectral indices are given according to $f_{\nu} \propto \nu^{\alpha}$ throughout, such that $\alpha$ represents the local slope of the SED in a log-log plot.

\section{Data}
\label{sec:data}

Our investigation is based on archival data described in Section~\ref{sec:archive} and new {\em Chandra} data described in Section~\ref{sec:newdata}.  Many of the selection criteria used are common to all samples and are summarized here.  In general, we have limited ourselves to quasars included in the \citet{srh+10} quasar catalog from the 7th Data Release (DR7; \citealt{aaa+09}) of the Sloan Digital Sky Survey (SDSS; \citealt{yaa+00}) such that we can make use of the SDSS spectroscopy of these sources.  The exception is for some PG/BQS \citep{sg83} quasars where there exists space-based UV spectra of the \civ\ region.  

We use the rest-frame \civ\ EQW and \civ\ blueshift as our main diagnostics.  We have chosen not to include the commonly-used \civ\ FWHM measurement in our analysis as that parameter is difficult to interpret in the disk+wind model of the \civ\ emission region that we have adopted in \citet{Richards2011}.  This concern is supported by the work of \citet{wwz+11}, who found that \civ\ has contributions from both an outflowing and a gravitationally bound component (the relative contributions of which may be dependent on the Eddington ratio).   \citet{sgs+08} showed that \civ\ FWHM provides a biased estimator of the black hole mass, but argued that this bias can be calibrated out in the ensemble average.

Except for the PG quasars, we matched the objects to the database that we used in \citet{Richards2011} and extracted our own values for the \civ\ line parameters and corrected SDSS redshifts as given by \citet{HW10}\footnote{Redshifts for all the quasars included in the Schneider et al.\ (2010)
DR7 quasar catalogue can be obtained from http://www.sdss.org/dr7/products/value\_added/index.html\#quasars .}.  BALQSOs and radio-loud (RL) quasars were excluded as the former are known to be absorbed in the X-ray \citep[e.g.,][]{gm96,gbc+02} and the latter are known to have jet-enhanced X-ray emission \citep[e.g.,][]{wtg+87,mbs+11}.  BALQSOs were taken from \citet{Allen10} and RL quasars were defined according to $\log L_{\rm 20\,cm} >32.0\; {\rm ergs\,s^{-1}\,cm^{-2}\,Hz}$ \citep[e.g.,][]{gkm+99}, or (for the PG quasars) $\log(R)>1$; see \citet{Richards2011}.

The following criteria were applied to limit the samples to objects with robust \civ\ measurements from SDSS and that are not significantly reddened (see \citealt{Richards2011} for more details): $z_{\rm em}>1.54$, EQW$_{CIV} >$ 5\AA (rest frame), $\alpha_{\rm UV} >-9$ (an error code), $\sigma_{\lambda CIV} <$ 10\AA , FWHM$_{CIV} >$ 1000, FWHM$_{CIV} > 2\sigma_{\rm FWHM_{CIV}}$, EQW$_{CIV} > 2\sigma_{EQW_{CIV}}$, $\Delta (g-i)\le0.3$ \citep{rhv+03}.  We further excluded objects 
with error codes as noted in the papers below.  
Finally, given the heterogeneous nature of the data set, we do not attempt to include objects with X-ray non-detections using partial correlation analysis \citep[e.g.,][]{ifn86}; however, this choice imposes a bias that should not be completely neglected.

Table~\ref{tab:tab1} lists all of the data sources, the number of objects matched to our database of SDSS-DR7 spectral properties, the number removed because they are BALs, RL, outside of the redshift range covered (\civ\ must be visible in the SDSS spectra), or have optical properties that do not meet the above criteria.  Objects can be rejected for more than one reason.
Thus, we also tabulate the final number of objects kept (after resolving duplicate entries [in favor of the most recent data] between the samples).  
Finally, we tabulate the redshift and $l_{\rm uv}$ range of the objects kept from each sample.  The objects from \citet{ssb+06} are not tabulated in Table~\ref{tab:tab1} as they are not matched to the SDSS database; see below.

\begin{figure}[h!]
\epsscale{0.75}
\plotone{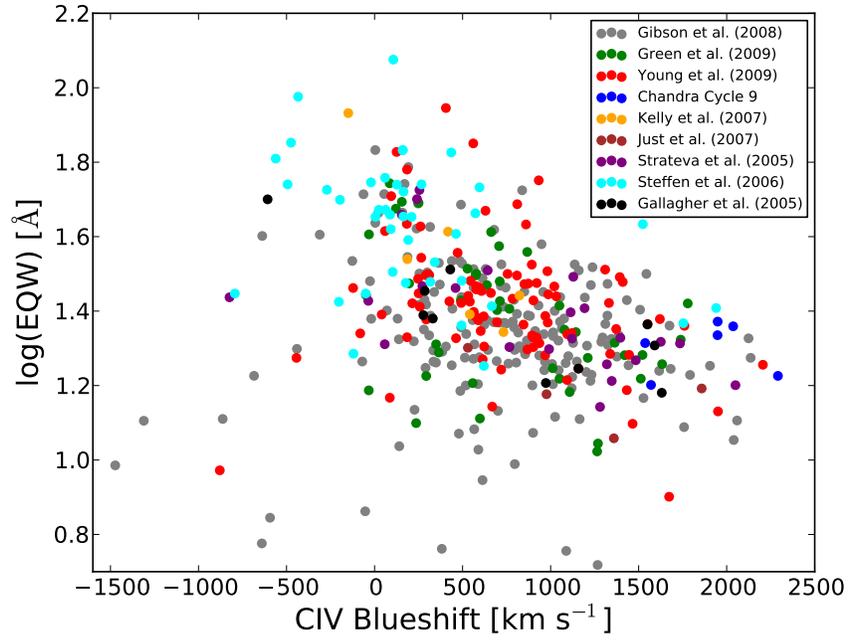}
\caption{Distribution of quasars in the \ion{C}{4} blueshift vs.\ log(EQW) parameter space. There are 409 objects in all; color-coding is by the source reference given in the legend.}
\label{fig:distribution}
\end{figure}

\begin{figure}[h!]
\epsscale{0.75}
\plotone{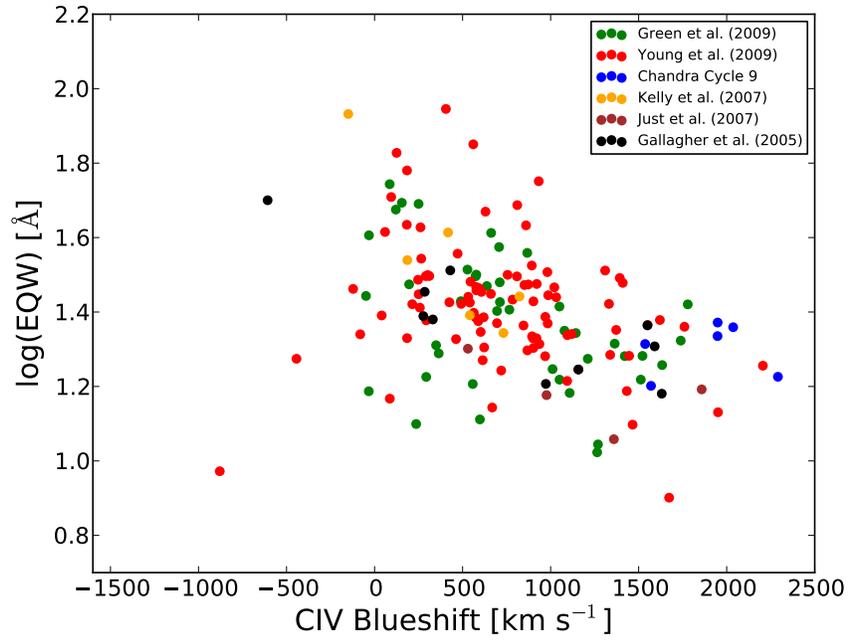}
\caption{As per Fig~\ref{fig:distribution}, but limited to sources with measured values of $\Gamma\equiv 1-\alpha_x$.  There are 164 objects in all.}
\label{fig:gamdistrb}
\end{figure}

In the end, we have a sample of 409 unique quasars.
Figure~\ref{fig:distribution} shows the distribution of the objects in \civ\ EQW-blueshift parameter space.
Of these objects, 164 have values for the X-ray photon index, $\Gamma$; their distribution is shown in Figure~\ref{fig:gamdistrb}.

\subsection{Archival Data}
\label{sec:archive}

In \citet{grh+05} we used {\em Chandra} to explore the X-ray properties of SDSS quasars at the extrema of the \civ\ blueshift distribution; archival observations filled in the blueshift distribution and provided additional objects at the extrema.  In the interim many more investigations of (largely) serendipitous quasar observations have been conducted by a number of groups (culminating in the {\em Chandra} Source Catalog\footnote{http://cxc.harvard.edu/csc} project), greatly simplifying the use of archival data.  As such, in addition to new {\em Chandra} observations described below, this investigation takes advantage of archival X-ray data from the following sources: \citet{grh+05}, \citet{sbs+05}, \citet{ssb+06}, \citet{kbs+07}, \citet{jbs+07}, \citet{gbs08}, \citet{gar+09}, and \citet{yer09}.   See \citet{wbh+11} for an investigation of objects even more extreme than considered herein.

The values taken from the following papers were $L_{2500}$, \aox, \daox, and $\Gamma$, where $\alpha_{ox} \equiv 0.384 \times \mbox{log}(f_{2 keV}/f_{2500})$ is the flux ratio between the X-ray portion of the SED (at $2\,{\rm keV}$) and the optical/UV portion (at $2500\,{\rm \AA}$) of the SED.  We will follow standard convention and abbreviate $\log L_{2500}$ as $l_{\rm uv}$.  \daox\ is the luminosity-corrected value of \aox, defined by $\Delta \alpha_{\rm ox} \equiv \alpha_{\rm ox} - \alpha_{\rm ox}(L_{2500})$ using the $L_{\rm UV}$--$\alpha_{\rm ox}$ relationship from \citet{jbs+07}.  Our sign convention is such that for both \aox\ and \daox, more negative values indicate (relatively) weaker X-ray sources.  $\Gamma$ is the X-ray photon index, which is related to the standard spectral index in the X-ray regime by $\Gamma\equiv1-\alpha_{\rm x}$.
If $L_{2500}$ was not given, we instead used $L_{2800}$ from our database.  For a power-law spectral index of $\alpha_{\rm opt}=-0.44$ \citep{vrb+01}, $\log L_{2500} = \log L_{2800}-0.02$ (and $\log L_{2500} = \log L_{1550}+0.09$).  While our 2800\,\AA\ luminosities are continuum luminosities and $L_{2500}$ typically includes emission-line flux, the difference is typically negligible when it comes to computing \aox\ (\aox\ changes by only $0.04$ for a 10\% line contribution).  If \daox\ was not given, we derived it from $L_{2500}$ and \aox.
Only some of the samples included measurements of $\Gamma$.    

\citet{sbs+05} cataloged 155
SDSS-DR2 quasars in {\em ROSAT} medium-deep fields.  
In all, 86\% were detected in the X-ray.  The {\em ROSAT} objects had an average exposure time of 16.7\,ks and were not targeted by SDSS solely because of their X-ray detections; only four {\em ROSAT} detections were from pointed observations of the quasars themselves.  From their Table 1 we extracted $L_{2500}$ and $\alpha_{\rm ox}$.  There is no information on $\Gamma$ for these data.  All other parameters (e.g., \civ\ and \daox) were determined as described above.   

\citet{ssb+06} extended the work of \citet{sbs+05} to lower luminosities using data from the COMBO-17 survey in the E-CDF-S \citep{wmk+04} and with 46 low-$z$ quasars from the PG sample.  As \citet{bl05} provide the \civ\ measurements for the PG sample, we were able to include the objects from the PG sample in our analysis; however, these are not matched to the SDSS spectral database and thus are not tabulated in Table~\ref{tab:tab1}.   Table~2 from \citet{ssb+06} provided values for $L_{\rm 2500}$ and \aox; no information on $\Gamma$ is available.
No SDSS \civ\ data exist for the COMBO-17 objects and they are not included here.  In addition, while \citet{ssb+06} also utilize 19 $z>4$ quasars, many of these are outside of the SDSS footprint --- as such, they lack other parameters needed for our analysis and they are also excluded.

\citet{kbs+07} report X-ray properties for seven new high-redshift quasars and 167 archival quasar with $z<4$; of these 157 match to our SDSS-DR7 spectral database.  The values for \aox, $\Gamma$, and $L_{2500}$ were extracted from their Table 4.  The energy range for $\Gamma$ was 0.3--7.0\,keV; most sources had insufficient counts to fit simultaneously for intrinsic $N_{\rm H}$.  Using these data, \citet{kbs+07} confirmed (at the $3.5\sigma$ level) the anti-correlation between $\Gamma$ and $\alpha_{\rm UV}$ seen by \citet{grh+05}.

\citet{jbs+07} examine 59 highly luminous quasars with $z>1.5$, including 21 objects with new {\em Chandra} observations; 32 objects are SDSS-DR3 quasars.
From their Table 4, we extracted $L_{2500}$, \aox, and \daox.  
$\Gamma$ values are taken from their Table 3; the energy range was 0.3--8.0\,keV and the signal-to-noise ratio was not high enough to justify additional constraints on $N_H$ for individual sources.  
For most sources no intrinsic absorption was allowed when fitting for $\Gamma$; however, sources with over 200 counts were simultaneously fit for $\Gamma$ and $N_H$.

\citet{gbs08} catalog {\em Chandra} and {\em XMM-Newton} observations of 536 SDSS-DR5 quasars covering $1.7<z<2.7$.  Most of the observations are serendipitous; less than 9\% of the quasars were the targets of the X-ray observations.  We obtained $L_{2500}$, $\alpha_{\rm ox}$ and \daox\ from their Table 1.  No X-ray spectral indices ($\Gamma$) are available.  Using these data \citet{gbs08} argue that the spread of \daox\ is mostly due to variability (see also \citealt{hgr+06}) and that the fraction of quasars that are intrinsically X-ray weak by a factor of 10 or more  is $<2$\%.  

\citet{gar+09} present {\em Chandra} data for 1135 spectroscopic and photometric SDSS-DR6 quasars analyzed by the Chandra Multiwavelength Project (ChaMP; \citealt[][]{gsc+04}); we include the fraction of these that have SDSS spectroscopic redshifts.  From their Table~2 we extract values for $\Gamma$, $L_{2500}$, and \aox\ (flipping the sign of \aox\ to agree with our convention).  For objects with more than 200 counts, $N_H$ was fit in addition to $\Gamma$; the energy range was 0.5--8.0\,keV.  

Finally, \citet{yer09} tabulate serendipitous {\em XMM-Newton} observations of 792 SDSS-DR5 quasars.  473 of these quasars had sufficient S/N to determine $\Gamma$ (and $N_H$, if warranted).  The spectral range for fitting was 0.5--10\,keV.  $\Gamma$ and \aox\ values are taken from their Table~2.

\subsection{New Data}
\label{sec:newdata}

While our goal is to understand the physics of quasars in the ensemble average, we often gain critical insight by first trying to understand the extrema.  This is as true for \civ\ blueshifts as it is for any other quasar parameter.  As outliers in any distribution are rare, we have expanded the sample of high-blueshift quasars with X-ray data by observing seven new radio-quiet sources in {\em Chandra's} Cycle 9.  We also obtained additional time on three sources from \citet{grh+05} to improve the S/N for those sources.  The new targets were chosen to be the highest blueshift quasars ($>1500{\rm \,km\,s}^{-1}$) from the SDSS-DR5 sample that met the following criteria.  Redshift was limited to $1.6<z<2.2$ such that both CIV and MgII are seen.  BALQSOs and RL quasars were excluded as discussed above.
Finally, to maximize our X-ray counts, we further limited the sample to $i<17.5$ and Galactic $E(B-V)<0.04$.  

The targets are summarized in Table~\ref{tab:chandra}.  Exposure times were set so as to achieve 100 total counts for each object.  All targets were detected with between 63 and 140 counts, and fluxes, luminosities, and photon indices were derived from the soft- and hard-band photometry following the methods described in \citet{gbc+06}.  Spectra with sufficient counts were analyzed with XSPEC \citep{XSPEC} using unbinned data and C-statistics to constrain $N_H$ and $\Gamma$.  The X-ray properties of the sample are summarized in Table~\ref{tab:xcalc} and these objects are combined with the archival data above in our analysis.

Three sources in the \citet{grh+05} sample, J0051$-$0102, J0147$+$0001, and J0208$+$0022, were reobserved in our new program with the goal of increasing the counts in the combined spectra for each object.  Between the observations, the sources showed significant flux variability of $-$52\%, $+$26\%, and $+$45\%, respectively.  While the $\Gamma_{\rm HR}$ values between epochs were consistent within the 1$\sigma$ error bars, the low counts in the individual spectra could obscure significant spectral variability as well.  Combined fitting of the new and old data together was inconclusive (in terms of the presence of absorption), likely because of the substantial variability between epochs.

\section{Analysis}
\label{sec:analysis}

The key results from \citet{Richards2011} and \cite{grh+05} were to provide a bridge between properties of the broad emission lines in quasars with their spectral energy distributions.  With this in mind, we further consider the X-ray properties of the heterogeneous data set described above.  We specifically consider the distributions of \aox, \daox, and \ax\ in the \civ\ EQW-blueshift parameter space as defined by \citet{Richards2011}.  If the differences seen across the range of \civ\ emission-line properties are indicative of a range of abilities to drive a strong accretion disk wind, and if radiation line driving plays a significant role in driving such a wind, then we might expect to see differences in X-ray properties of quasars across the \civ\ EQW-blueshift parameter space.  Ultimately, our hope is that this work is a first step toward improving the effective resolution of quasar SEDs in the ``unseen'' part of the continuum that covers nearly two decades in frequency between the optical and the X-ray.

\subsection{$\alpha_{\rm ox}$}

In a disk-wind picture, quasars with radiation-pressure dominated winds should be {\em intrinsically} X-ray weak relative to the optical/UV and are likely to display X-ray warm absorbers \citep{grh+05} that may {\em additionally} reduce the soft X-ray flux along our line of sight.  Our working hypothesis is that an object's location in the \civ\ EQW-blueshift parameter space reflects its intrinsic ability to drive a strong accretion disk wind.  Assuming that radiation line driving dominates the wind component, we expect to see a reduction in \aox\ across the \civ\ EQW-blueshift parameter space (from top left to bottom right in Figure~\ref{fig:distribution}) as the hypothetical wind strength increases.  See \citet[Figure~7]{lm04}, for evidence of this behavior in NLS1s.

\begin{figure}[h!]
\epsscale{0.7}
\plotone{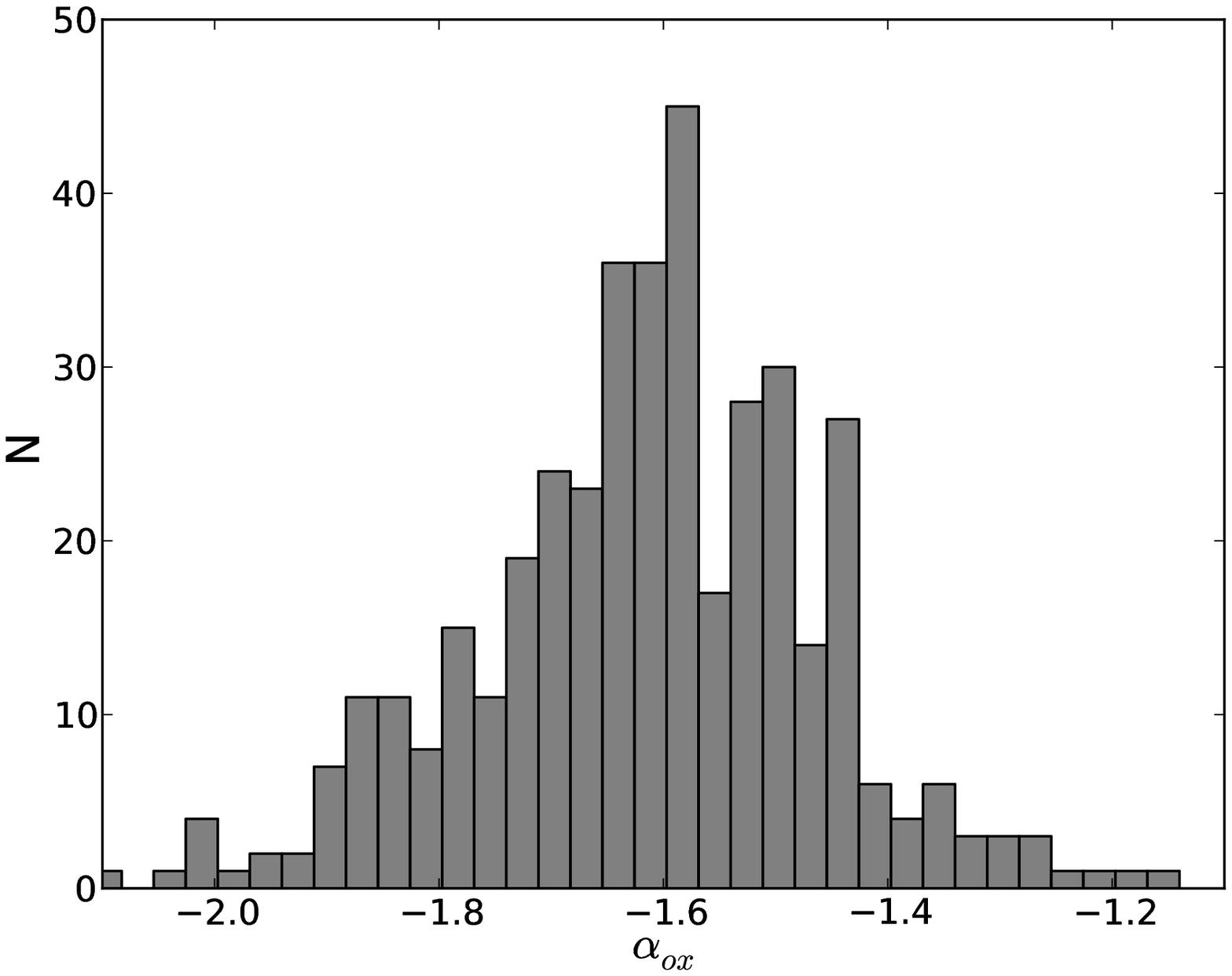}
\caption{Histogram of \aox\ values in the sample and used in Figure~\ref{fig:aox}.    The mean and standard deviation are $-1.619\pm0.143$.
}
\label{fig:aoxhist}
\end{figure}

Figure~\ref{fig:aoxhist} gives the range of \aox\ values in the sample, while
Figure~\ref{fig:aox} shows how \aox\ changes across the \civ\ EQW-blueshift parameter space using two different diagnostics.  First we show the location of the individual quasars in our heterogeneous sample color-coded by \aox\ (circles) as indicated by the color-bar on the right-hand side.  The dashed lines indicate the divisions used in \citet{Richards2011} to create composite spectra as a function of \civ\ emission-line properties (see also Section~\ref{sec:seds}).  There appears to be a general trend towards more negative \aox\ values (objects relatively weaker in the X-ray) from the top-left to the lower-right.  However, as there is considerable scatter in the diagram, we also bin the results to better discern the general trend.  Specifically the colored squares indicate the median values of \civ\ EQW and blueshift after sorting the values of \aox\ and grouping them into 10 bins.  
Although the median values span a much smaller range in \civ\ properties than the raw sample, they clearly reveal a general trend towards X-ray weaker objects in the lower right-hand corner.

Further evidence for a trend in \aox\ with \civ\ properties comes from performing a Student's $t$-test for different means \citep[e.g.,][]{ptv+92} using the objects in the upper-left and lower-right regions of the \civ\ parameter space in Figure~\ref{fig:aox}.  We find that the mean values for \aox\ in those bins are $-1.497$ and $-1.700$, respectively, with a standard deviation of $0.021$, yielding a $t$ value of $9.85$ with very high significance that the samples have different means.

This trend of decreasing relative X-ray strength as we move from the upper-left (large EQW, small blueshift) to the lower-right 
(small EQW, large blueshift) in Figure~\ref{fig:aox} is consistent with the expectations of a model of the BELR where a radiation line driven wind is influencing the properties of \civ.  When the X-ray flux is high, a strong radiation line driven wind cannot form (though perhaps there is still an MHD-driven wind; \citealt{Proga03}) and the ionizing continuum photons can create a large population of triply ionized carbon atoms in the outer part of the accretion disk.  If instead, the X-ray flux is weaker, a strong radiation line driven wind can form, shielding the outer accretion disk from an already weaker X-ray flux.  In this case the \civ\ line is formed mostly in the wind itself and may be blueshifted due to the outflowing nature of the source and lower in equivalent width due to the overall reduction in ionizing photons \citep{Leighly04}.  It is less clear what is happening in the lower-left corner (small EQW, and small blueshifts), but it would appear that those objects have more heterogeneous \aox\ values than the objects in the upper-left or lower-right corners.  Third parameters, such as covering fraction, orientation, absorption, and/or variability effects could explain objects in the lower left-hand corner (\citealt[e.g.,][]{hc10}; Bowler, Allen, \& Hewett, in preparation).  As noted in \citet{Richards2011}, objects with both large \civ\ blueshift and large \civ\ EQW do not seem to exist and this is unlikely to be a selection effect.

Our results are consistent with the results of \citet{wvb+09} who have done a multivariate regression of EQW against both $l_{\rm uv}$ and \aox.  They find that \civ\ EQW correlates with \aox\ such that harder spectra have stronger \civ.  What our work does is to further separate low EQW objects in another dimension.   While there is no degeneracy for large EQW quasars, those with smaller \civ\ EQW values can have a range of \civ\ blueshifts.  Since \aox\ is also correlated with \civ\ blueshift (in a way that is not trivially related to the \civ\ EQW), it is important to explore trends in \aox\ in this 2-D parameter space.

\subsection{$\Delta \alpha_{ox}$}

As it is well known that \aox\ is correlated with UV luminosity \citep[e.g.,][]{at82,jbs+07}, the above trends with \aox\ in the previous section might instead be ascribed to luminosity.  Indeed, we find that UV luminosity is increasing from the upper-left to the lower-right in Figure~1.  As such, it has become common to reference \aox\ values to the mean \aox\ values observed for objects with the same UV luminosity according to $\Delta \alpha_{ox} \equiv \alpha_{ox} - \alpha_{ox}(L_{2500})$.  Here we adopt the expected value of \aox\ from \citet{jbs+07}, specifically $\alpha_{ox}(L_{2500}) = (-0.140 \pm 0.007) \times \mbox{log}(L_{2500}) + (2.705 \pm 0.212)$.  $\Delta \alpha_{ox}$ indicates the relative amount of X-ray emission, whereby negative values of $\Delta \alpha_{ox}$ indicate a relative deficit of X-rays, while positive values indicate an X-ray surplus.   Figure~\ref{fig:daoxhist} shows the distribution of \daox\ values in our sample.

By using \daox\ instead of \aox, luminosity is marginalized and any remaining SED effects must instead be due to the {\em shape} of the SED rather than its absolute value.  It is therefore significant that Figure~\ref{fig:daox} shows the same general trend as does Figure~\ref{fig:aox}.  While taking out the luminosity dependence has apparently reduced the trend for individual objects, the median values of \civ\ EQW and blueshift (using 10 bins sorted in \daox) again show a significant trend from more X-ray luminous objects in the upper left-hand corner to relatively weaker X-ray sources in the lower right-hand corner. 
\citet{gbs08} similarly find that low EQW objects and large blueshift objects have more negative values of \daox\ (their Figures 8 and 10), although it should be noted that half of our sample comes from their paper, so our results are not independent.
As with \aox, a Student's $t$-test confirms that the mean values of \daox\ in the upper-left and lower-right parts of the \civ\ distribution in Figure~\ref{fig:daox} are significantly different.   We find respective mean values of $0.103$ and $-0.030$ with a standard deviation of $0.021$, yielding $t=6.55$ with high significance.

Unfortunately our understanding of the X-ray SED data is in roughly the same state as our understanding of the optical/UV SED was in the 1990s.  \citet{fhf+91} were able to establish that the optical/UV spectral index was $\sim-0.4$ with an error of $\sigma_{\alpha}\sim0.5$; however, it wasn't until \citet{rhv+03} that the photometric errors were small enough that the $\alpha_{\rm uv}$ distribution could be shown to have an intrinsic spread rather than a single characteristic value smeared by measurement errors.  In the case of the X-ray, variability likely plays more of a role than measurement errors especially since both the optical and X-ray measurements are subject to changes over time and are non-simultaneous.  Indeed, the majority of the width in the \daox\ distribution can be attributed to variability \citep{hgr+06,gbs08}.  The width of the \daox\ distribution notwithstanding, there does appear to be a trend towards relatively weaker X-ray sources in objects with large \civ\ blueshifts.


\subsection{$\Gamma$}

While the \aox\ and \daox\ trends with \civ\ EQW and blueshift are expected in a radiation line driven wind model, we have less intuition regarding how the X-ray spectral indices ($\alpha_{\rm x} = 1- \Gamma$) should behave.  At low redshift, where X-ray data sample the soft X-ray part of the spectrum, it is observed that $\Gamma_{\rm soft}$ correlates with the strength of \ion{Fe}{2} (relative to H$\beta$) and anti-correlates with the FWHM of H$\beta$ \citep{szm+00}; X-ray softer objects thus have stronger \ion{Fe}{2} and narrower H$\beta$ \citep{lfe+97}.  In the context of black hole binaries, such soft-spectrum objects (including NLS1s; \citealt{bbf96}) are thought to have high accretion rates \citep{pdo95}.
Unfortunately there are relatively few objects with spectral coverage of both the H$\beta$ and \civ\ line, so it is hard to know how these trends would propagate to higher redshift objects.  However, using {\em Hubble Space Telescope} data on \civ\ for low-$z$ quasars, \citet{sbm+07} have shown that large blueshift objects tend towards softer $\Gamma_{\rm soft}$.   The $\Gamma$ values tabulated here (Fig.~\ref{fig:gammahist}) generally refer to the harder part of the X-ray spectrum from 2\,keV to 10\,keV, so it does not necessarily follow that large blueshift objects should tend towards softer $\Gamma_{\rm hard}$; however, that does seem to be the case.  If higher UV luminosity results in more Compton cooling of the corona, then one indeed might expect $\Gamma_{\rm hard}$ to follow $\Gamma_{\rm soft}$ \citep{pdo95,lfe+97} and that quasars with large blueshift might have high accretion rates.  

Figure~\ref{fig:gamma} shows a weak trend of increasing (softer) $\Gamma$ with increasing blueshift and decreasing \civ\ EQW.   As with the \aox\ and \daox\ figures the median \civ\ properties as a function of $\Gamma$ show a clearer trend than can be discerned from looking at the distribution of individual objects.  
We find a Student's $t$ of $-3.221$ with significance of $0.002$, showing that there is a clear difference when comparing the means of objects in the top-left ($\Gamma=1.829$) and the lower-right ($\Gamma=2.052$) parts of Figure~\ref{fig:gamma}; the standard deviation is $0.069$.  Clearly further work is needed to better understand whether the the harder $\Gamma$ values for strong \civ\ quasars are indicative of the intrinsic X-ray spectrum.

\section{Discussion}
\label{sec:discussion}

\subsection{Empirical SEDs}
\label{sec:seds}

The values of \aox, \daox, and $\Gamma$ are perhaps more meaningful within the context of the overall SED.  As such, in Figure~\ref{fig:sed} we create typical SEDs spanning the \civ\ EQW-blueshift bins used by \citet{Richards2011}; the \civ\ properties of each SED are shown in Figure~\ref{fig:sedpos}.  
For this analysis, we imposed a $z<2.2$ limit such that both \civ\ and \ion{Mg}{2} are observed and a UV spectral index between the continua underlying those lines could be determined empirically.  The SEDs were created using the {\em median} values of L$_{1550}$, $\alpha_{\rm opt}$, $\alpha_{\rm ox}$ and $\Gamma$ for each sub-sample, where the first two quantities were taken from our own work and the last two quantities were taken from the literature as described above (with the exception of our new {\em Chandra} data).  To make the SEDs, the median value of $\log(L_{1550}$) was used with the median $\alpha_{\rm opt}$ to calculate a value for log(L$_{\nu}$) at 2500\,\AA.  In turn, log(L$_{2500}$) was used with $\alpha_{\rm ox}$ to determine a value for log(L$_{2\, \rm keV}$).  Then $\alpha_{\rm x}$ was used to plot a point at $L_{10\,\rm keV}$.  In addition to our optical and X-ray results, we have included the UV results of \citet{Scott2004} who found that the UV spectral index over $\sim$650--1100\,\AA\ is a function of luminosity.  Specifically, we have used the results of Table~4 and Figure~18 from \citet{Scott2004} and our value of $L_{1550}$ (as a proxy for $L_{1100}$ used by \citealt{Scott2004}) to estimate the spectral slope over $500<\lambda ({\rm \AA})<1216$ [$15.78>\log({\rm freq [Hz]})>15.39$].  The luminosity dependences of $\alpha_{\rm UV}$ and \aox\ appear to be fairly comparable as the break at 500\,\AA\ is difficult to discern, which is consistent with the findings of \citet{lfe+97}.   

We emphasize that there is little information on the shape of the SED between  500\,\AA\ and $\sim0.2$\,keV for luminous quasars; \aox\ only parametrizes the UV to X-ray flux ratio, not the actual shape of the SED, which has implications for bolometric corrections; see Section~\ref{sec:bc}.
Interestingly, those objects that are relatively stronger at 2\,keV relative to 2500\,{\rm \AA} (i.e., have a harder ionizing spectrum) are also apparently harder (i.e., flatter) over 2--10\,keV.

Although the differences between the SEDs are subtle, the effects are clearly systematic.  Relative to their 2500\,\AA\ luminosities, the large \civ\ EQW objects have ``harder'' SEDs than the large \civ\ blueshift objects.  We illustrate this more clearly in the inset of Figure~\ref{fig:sed} by normalizing all of the SEDs to the same 2500\,\AA\ luminosity.  It is important to realize that while the \aox\ distribution for these SDSS quasars nowhere near spans the extrema as characterized by the weak-lined quasar PHL 1811 ($\alpha_{\rm ox}=-2.3$; \citealt{lhj+07}) and narrow-line Seyfert 1 galaxy RE 1034+39 ($\alpha_{\rm ox}=-1.2$; \citealt{clb06}), nevertheless, for the same UV luminosity, a change in \aox\ of just $0.2$ corresponds to a change in flux at 2\,keV of a factor of 3.  This difference is simply a reflection of the large lever arm between 2500\,\AA\ and 2\,keV (2.6 decades in frequency).  As such it is not unreasonable to expect significant differences in the ability of quasars to drive a wind through radiation line pressure as we move across the \civ\ EQW-blueshift parameter space.


\subsection{Comparison with Model SEDs}

Since there are few {\em empirical} constraints on the 500\,\AA\ to 0.2\,keV part of the SED, it is useful to compare our empirical SEDs to some example model SEDs.  We specifically consider model SEDs similar to those described in \citet{clb06} in accordance with the AGN spectrum in the CLOUDY package \citep{fer03}.  The optical/UV portion of the SED is modeled as a power law with exponential cutoffs at high and low energies as parametrized by a cutoff temperature (the temperature of the inner edge of the accretion disk for the high energy cutoff).  The optical spectral index is taken to be $\alpha_{\rm opt}=-0.33$ and the X-ray spectral index is fixed to $\alpha_{\rm x}=-1\; (\Gamma = 2)$.  The optical/UV part of the spectrum is related to the X-ray part of the spectrum by the well-known $L_{\rm UV}$--\aox\ relationship \citep[e.g.,][]{jbs+07}.  Note that, in this model, there is a degeneracy between $L_{\rm disk}$, $L_{\rm UV}$, $L_{\rm X}$, and $T_{\rm cut}$ that means that the SEDs are uniquely described by a single parameter that can be characterized by \aox.
In Figure~\ref{fig:bb} we show a model SED (solid black line) that best represents the average quasar in our sample.  Two examples that are meant to bracket the range of SEDs spanned by our \civ\ sample (in terms of $L_{\rm UV}$ and \aox) are also shown.
The two extreme SEDs shown have cutoff temperatures of 20\,eV and 50\,eV and have, respectively, $\alpha_{\rm ox} = -1.74$ and $\alpha_{\rm ox} = -1.5$, and 
$l_{\rm UV}$ of 31.8 and 30.0. 

As noted by \citet{lfe+97}, such model SEDs with very ``hot'' accretion disks are apparently inconsistent with the UV part of empirical SEDs \citep[e.g.,][]{ewm+94,rls+06}.  Indeed, the extreme models shown in Figure~\ref{fig:bb} have spectral indices of $-1.24$ and $-0.69$ at 800\,\AA, respectively, which are within the range observed by \citet{Scott2004}, but are too blue/hard for these luminosities as can be seen by comparison with our empirical SEDs (dotted lines).  However, there is evidence that the SED the emission-line gas sees may not be the same as the SED that we see \citep{kfb97} and our work herein and in \citet{Richards2011} might be taken as further evidence of such a situation.  If both the observer and the emitting gas {\em do} see the same SED then, if this model parametrization is accurate, our empirical data would suggest that a lower cutoff temperature might be more appropriate for the luminous quasars in our sample.  We are currently pursuing a more detailed characterization of the near-UV part of the spectrum for SDSS quasars (Krawczyk et al.\ 2011, in preparation).

Aside from the near-UV slope mismatch, the 20\,eV peaked spectrum is representative of the most luminous quasars in our sample (in terms of \aox).  The relatively soft spectrum (X-ray weak relative to UV) would be conducive to driving a strong wind through radiation line driving, yielding a large \civ\ blueshift.  On the other hand, the 50\,eV peaked spectrum is more representative of the less luminous quasars in our sample.  For such quasars, the relatively hard spectrum is more likely to overionize the accretion disk atmosphere and has less UV flux, both of which would tend to inhibit a radiation line driven wind.  This harder SED is one that we associate with quasars that have strong \civ\ emission that is dominated by the disk component of the BELR.

Although there is a lack of data to characterize the shape of the SED over much of the crucial ionizing part of the spectrum, our model SEDs represent extrema characterizing the ionizing SED for a given \aox\ (c.f., the simple power-law model, dotted lines in Figure~\ref{fig:bb}) in that they have as much far-UV flux as one could imagine.  Thus, these extrema are particularly interesting to consider in terms of bolometric corrections; see Section~\ref{sec:bc}.  An intermediate model might be one that has a cooler accretion disk with a Comptonized extension to soft X-ray energies \citep[e.g.,][]{lfe+97,gd04}, with another, hotter Comptonized region producing the hard X-ray spectrum.  For example, even if the accretion disk is not as hot as depicted by the models in Figure~\ref{fig:bb}, \citet{kfb97} suggest that quasars with strong \ion{He}{2} (which tend to have strong \civ\ at the systemic redshift; \citealt{Richards2011}) would require a factor of a few higher flux at $\sim50$\,eV than most empirical SEDs (including ours) would suggest.

Comparison of the shape of the model SEDs in Figure~\ref{fig:bb} is relevant not only for bolometric corrections, but may also relate to the origin
of soft X-ray excesses \citep[e.g.,][]{abc+85,wf93,gd04,cfg+06,dc07,sd07}.  Specifically, since the UV through X-ray SEDs of radio-quiet quasars appear to change significantly from one corner of \civ\ parameter space to the other, it is interesting to ask where soft X-ray excess objects lie in the \civ\ EQW-blueshift parameter space.  Unfortunately the bandpasses of {\em Chandra} and {\em XMM-Newton} are such that it is difficult to identify soft X-ray excesses in high-redshift quasars.  However, the 21 low-redshift PG quasars studied by \citet{pro+04} have UV measurements of \civ\ from \citet{bl04}, \citet{bl05}, and \citet{sbm+07} and can be placed in our \civ\ parameter space. 
We find that the \citet{pro+04} objects are generally in the upper-left of the \civ\ parameter space (i.e., they tend to have large EQW and small blueshift, which, ironically, suggest that soft excess objects are actually X-ray hard [relative to the UV]).  Although the \citet{pro+04} objects are much less luminous than the average quasars in our sample ($l_{\rm uv}\sim29.7$), they are actually well matched to the lowest luminosity (and generally hard-spectrum) objects of our sample (being bright objects at low redshift as opposed to faint objects at high redshift).  
As such, it would be valuable to explore the possibility of correlations between the \civ\ properties of quasars and the presence of a soft X-ray excess.
We specifically predict that the average high-luminosity, high-$z$ quasar does not have an X-ray soft excess based on the relative locations in the 2D \civ\ parameter space of low-$z$ soft excess objects.

Lastly, we note that
a strict linear interpolation between 2500\,\AA\ and 2\,keV would predict a {\em significant} soft X-ray excess in all of our sources (compare the empirical and model SEDs in Figure~\ref{fig:bb} for energies higher than 0.2\,keV).  A lack of such a strong soft X-ray excess in luminous quasars would mean that while \aox\ may accurately indicate the 2\,keV flux density in quasars, it is a poor characterization of the shape of the true ionizing SED over $\sim10$--1000\,eV and we reiterate that the models shown in Figure~\ref{fig:bb} essentially represent the opposite extrema.  Here we close by merely pointing out that the apparent hard-spectrum nature of disk-dominated BELRs and soft-spectrum nature of wind-dominated BELRs suggests that these two extrema may have different soft X-ray excess properties and that the presence or lack of a soft X-ray excess might provide a way to reverse engineer the detailed shape of the ionizing SED (which is important to radiation line driving).

\subsection{Bolometric Corrections}
\label{sec:bc}

AGN luminosities are often derived from observations in one, or at best a few, narrow bandpasses, whereas one wants the luminosity integrated
over the full SED (or perhaps just the accretion disk's contribution; \citealt{mrg+04}), i.e., the bolometric luminosity.  For example, the bolometric luminosity is the relevant quantity needed to estimate the accretion rate through $L=\eta\dot{M}c^2$.  

Considering our empirical SEDs and the range of \aox\ values in our sample, we can see from Figure~\ref{fig:sed} that while the difference in SEDs may not appear to be very large, it is clearly {\em systematic}.  If we assume the same bolometric correction for all quasars --- regardless of their SEDs, then there will be a systematic error in $L_{\rm bol}$ as a function of their location in \civ\ EQW-blueshift parameter space.  Indeed, a change of 0.2 in \aox\ corresponds to a 35\%
 change in the integrated luminosity between 2500\,\AA\ and 2\,keV (assuming that the SED is described only by \aox).

If there are additional features in the SED at far-UV wavelengths \citep[e.g.,][]{Scott2004} or if the shape of the EUV spectrum is more like the models in Figure~\ref{fig:bb} than our power-law fits, then that is an additional source of systematic error in the bolometric correction that arises through the adoption of a single mean SED (or even one parameterized by the $L_{\rm UV}$--$\alpha_{\rm ox}$ relation). 
Indeed, \citet{rls+06} found that the there is no single bolometric correction, but rather a roughly factor of four range around the mean value.  While \citet{rls+06} were unable to determine which classes of quasars should use lower/higher bolometric corrections, our work here would suggest that bolometric corrections may be systematically underestimated for large EQW-small blueshift quasars (i.e., those with disk-dominated BELRs) and systematically overestimated for small EQW-large blueshift quasars (i.e, those with wind-dominated BELRs).  It is beyond the scope of this paper to specifically parameterize the bolometric correction as a function of the indirectly measured ionizing flux in the EUV; however, we hope to pursue that in a future publication (Krawczyk et al.\ 2011, in preparation).  Given the trends with UV luminosity in the EQW-blueshift parameter space, we might expect that more accurate bolometric corrections would result in a narrower distribution of $L_{\rm bol}$.  As $L_{\rm bol}$ is generally used to estimate the Eddington ratio (i.e., mass weighted accretion rate), $L_{\rm bol}/M_{\rm virial}$, those values may also have systematic errors.






\subsection{Nomenclature}

Given our findings on the systematic changes in SED over \civ\ EQW-blueshift parameter space, we consider the issue of nomenclature for these differences.

In \citet{Richards2011} we referred to the RQ quasars with small blueshifts and large EQWs as ``RL-like'' because RL quasars are predominantly found in that part of \civ\ parameter space.  
Based on our results above, we suggest a simpler, yet more descriptive nomenclature.  As RQ quasars in the RL part of parameter space have harder (more ionizing) spectra, such objects could be broadly categorized as hard-spectrum quasars (HSQs).  The high-blueshift, low-EQW quasars on the other hand have softer (less ionizing) spectra and could be considered to be soft-spectrum quasars (SSQs).  These distinctions are not dissimilar from the Population A/B terminology used by \citet{szm+00}, where the most extreme Population A objects would correspond roughly to SSQ and Population B to the HSQ objects.  The hard/soft terminology simply has the advantage of being physically motivated.

It is less clear what is happening in the low-EQW, low-blueshift quasars (occupying the other part of the ``Population A'' parameter space).  Weak lines 
could indicate 
a relatively weak ionizing continuum, but the lack of a strong blueshift might indicate that the ionizing continuum must be intrinsically weak rather than \aox\ being soft.  Another possibility is that \aox\ could be soft, but the UV flux is insufficient to drive a wind.  Further {\em Chandra} and/or {\em XMM-Newton} observations are needed to better understand the differences between sources with weak \civ\ that is systemic and weak \civ\ that is highly blueshifted.

If the winds in high-blueshift, low-EQW quasars (i.e., SSQs) are predominantly radiatively driven, their soft-spectrum nature cannot be generally due
to absorption along our line of sight, despite the fact that these objects appear to be the parent sample of BALQSOs.  That is because the SED needs to be {\em intrinsically} soft in order to drive a strong radiation pressure driven wind.  (MHD driving could also be at work, and may be a source of scatter in our diagrams.)  Rather, it must be that for BALQSOs, there is additional absorption of an SED that already intrinsically lies at the soft (X-ray weaker) extrema of the range of normal SEDs spanned by quasars
\citep[e.g.,][]{gm96,gbc+02}.

\section{Conclusions}
\label{sec:conclusions}

\begin{figure}[h!]
\epsscale{0.7}
\plotone{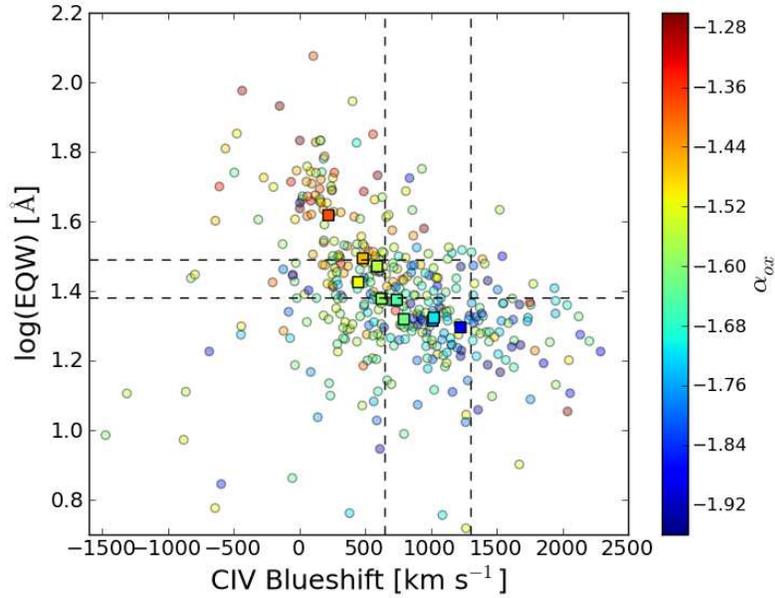}
\caption{\ion{C}{4} blueshift vs.\ log(EQW) colored by \aox\ values (transparent filled circles) as indicated by the color bar on the right.  Also shown are the median values of  \ion{C}{4} blueshift and log(EQW$_{CIV}$) after sorting the \aox\ values and dividing them into 10 bins (colored squares).  Objects with large \civ\ blueshift and small EQW are seen to have relatively X-ray weak SEDs.  Dashed lines refer to the eight regions used in Section~\ref{sec:seds} to make composite SEDs.
}
\label{fig:aox}
\end{figure}

\begin{figure}[h!]
\epsscale{0.7}
\plotone{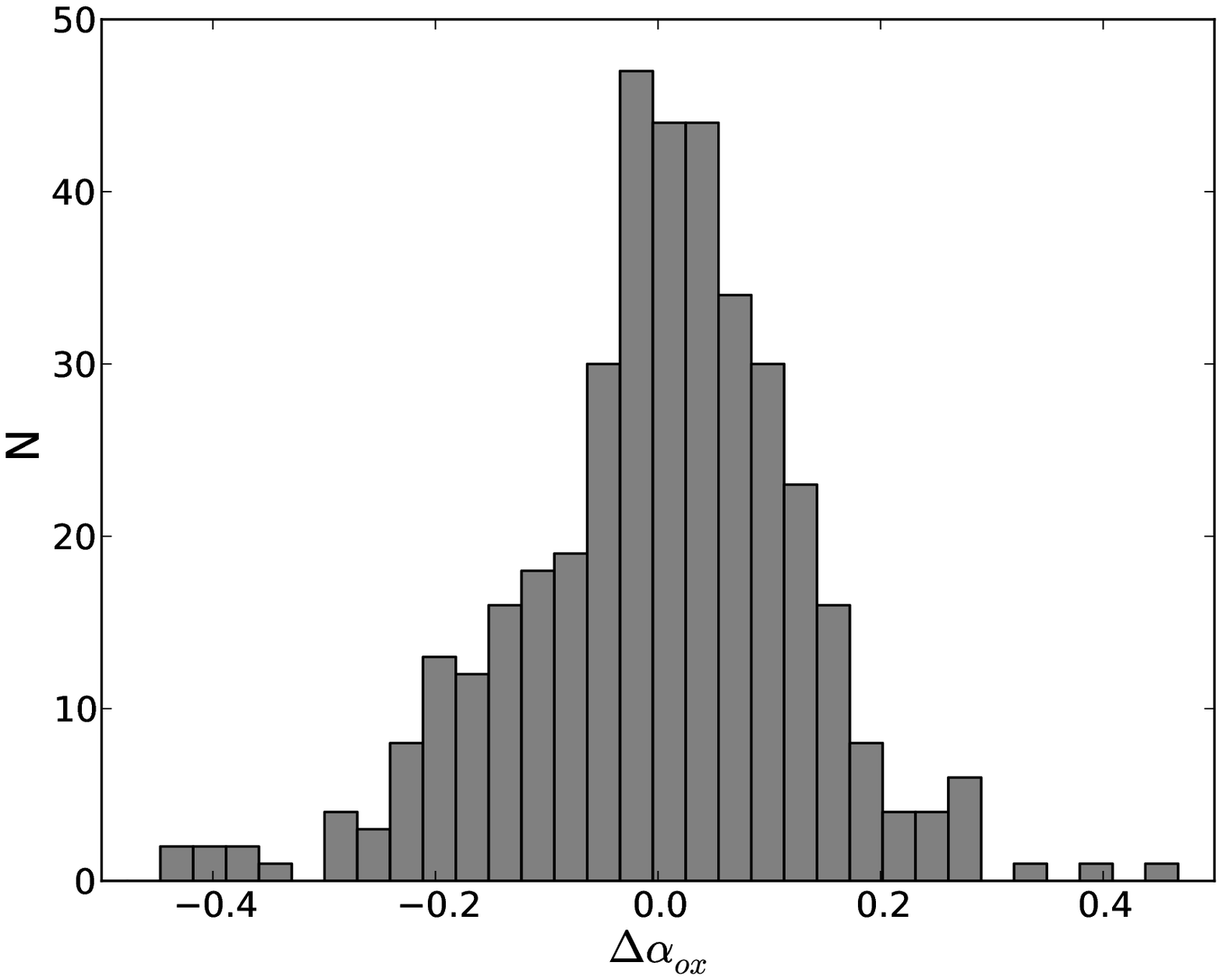}
\caption{Histogram of \daox\ values in the sample used in Figure~\ref{fig:daox}.   The mean and standard deviation are $0.008\pm0.126$.
}
\label{fig:daoxhist}
\end{figure}

\begin{figure}[h!]
\epsscale{0.7}
\plotone{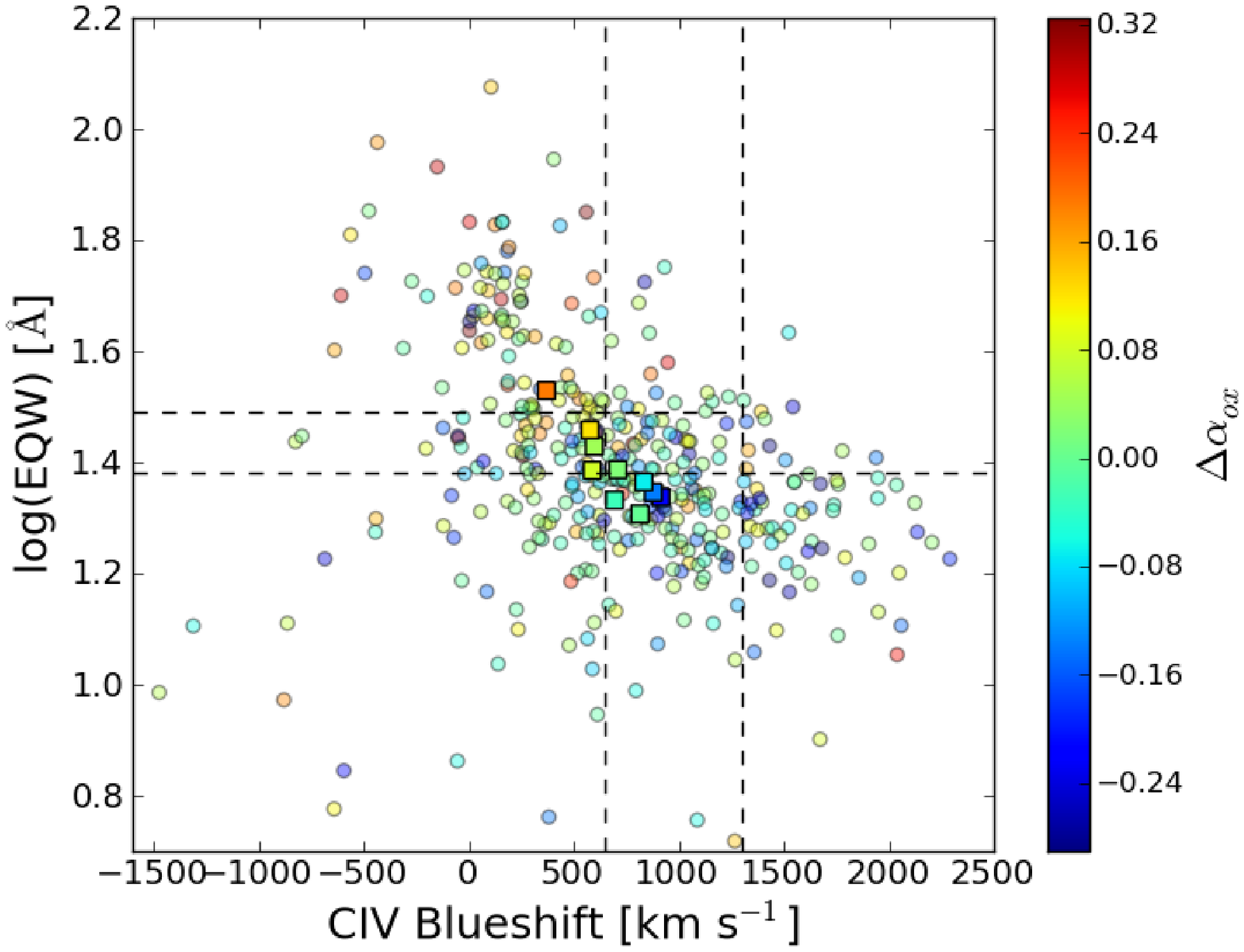}
\caption{\ion{C}{4} Blueshift vs.\ log(EQW) colored by \daox\ values (transparent filled circles) as indicated by the color bar on the right.  Also shown are the median values of  \ion{C}{4} Blueshift and log(EQW$_{CIV}$) after sorting the \daox\ values and dividing them into 10 bins (colored squares).  Thus even accounting for luminosity effects, objects with large \civ\ blueshift and small EQW are seen to have relatively X-ray weak SEDs.  Dashed lines refer to the eight regions used in Section~\ref{sec:seds} to make composite SEDs.
}
\label{fig:daox}
\end{figure}

\begin{figure}[h!]
\epsscale{0.7}
\plotone{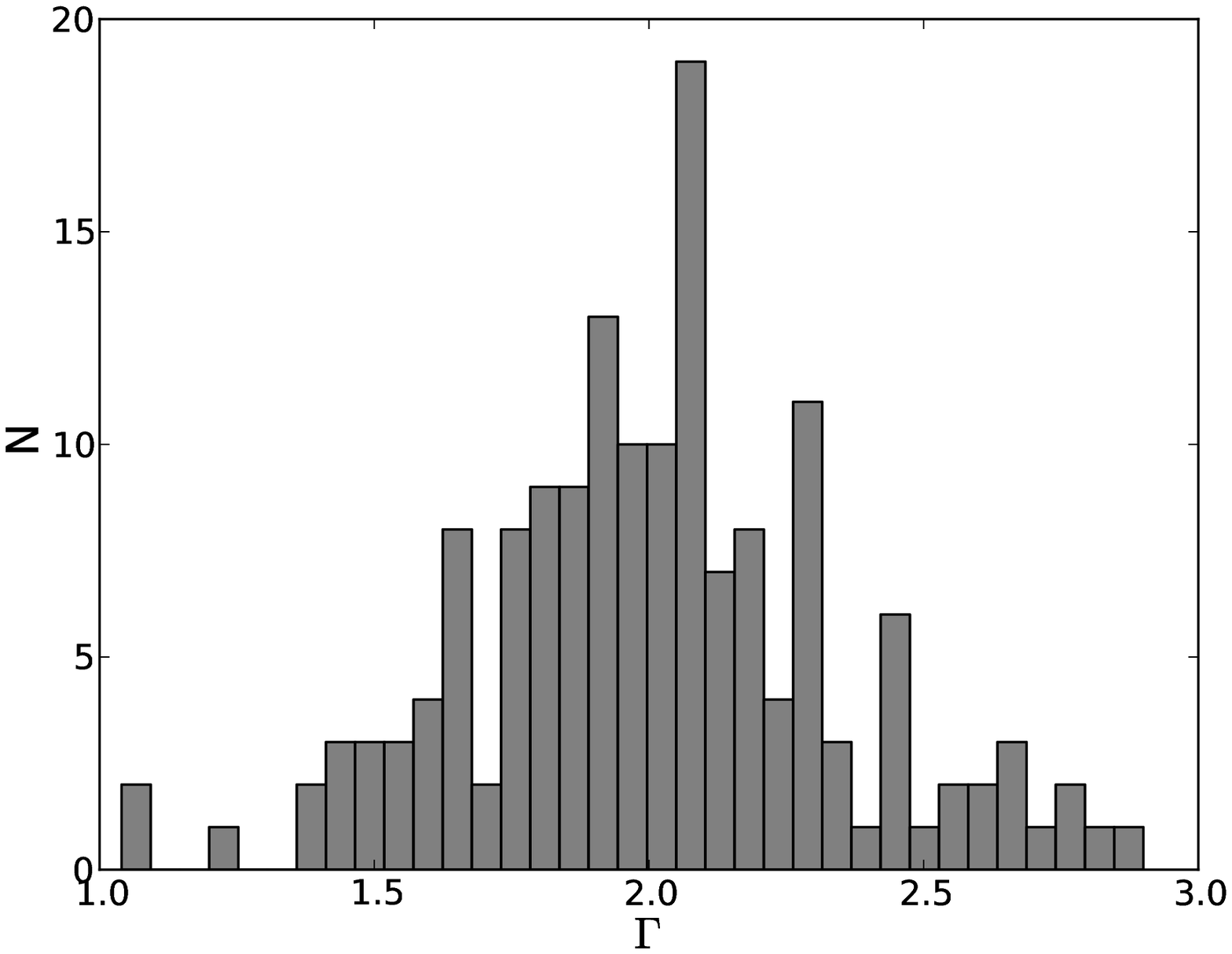}
\caption{Histogram of $\Gamma$ values in the sample used in Figure~\ref{fig:gamma}.   The mean and standard deviation are $2.005\pm0.326$.
}
\label{fig:gammahist}
\end{figure}

\begin{figure}[h!]
\epsscale{0.7}
\plotone{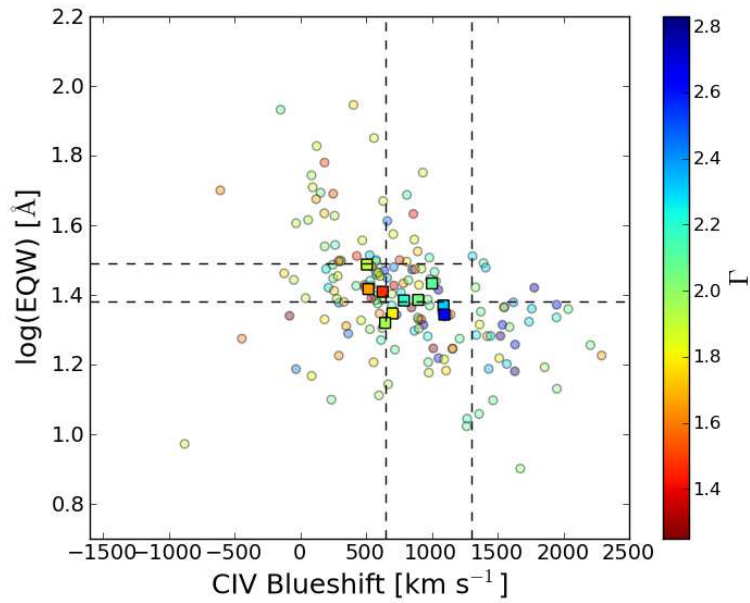}
\caption{
\ion{C}{4} Blueshift vs.\ log(EQW) colored by $\Gamma$ values (transparent filled circles) as indicated by the color bar on the right.  Also shown are the median values of  \ion{C}{4} Blueshift and log(EQW) after sorting the $\Gamma$ values and dividing them into 10 bins (colored squares).  Dashed lines refer to the eight regions used in Section~\ref{sec:seds} to make composite SEDs.
}
\label{fig:gamma}
\end{figure}

\begin{figure}[h!]
\plotone{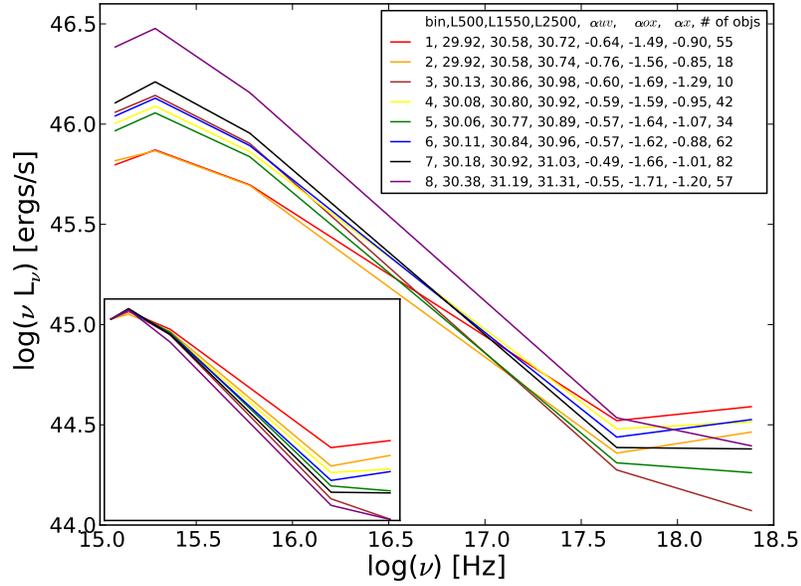}
\caption{SEDs for each of the eight \civ\ EQW-blueshift bins defined in Fig.~\ref{fig:sedpos}. The legend shows the bin number and plotting color each line corresponds to.  We also tabulate the median values for $L_{1550 \rm \AA}$, $\alpha_{\rm uv}$, $\alpha_{\rm ox}$ and $\alpha_{\rm x}$, the calculated value of $L_{2500\,{\rm \AA}}$ and the number of objects in each bin. (Inset:) SEDs normalized to the same $L_{2500\,{\rm \AA}}$ to show the differences in ionizing flux.}
\label{fig:sed}
\end{figure}

\begin{figure}[h!]
\plotone{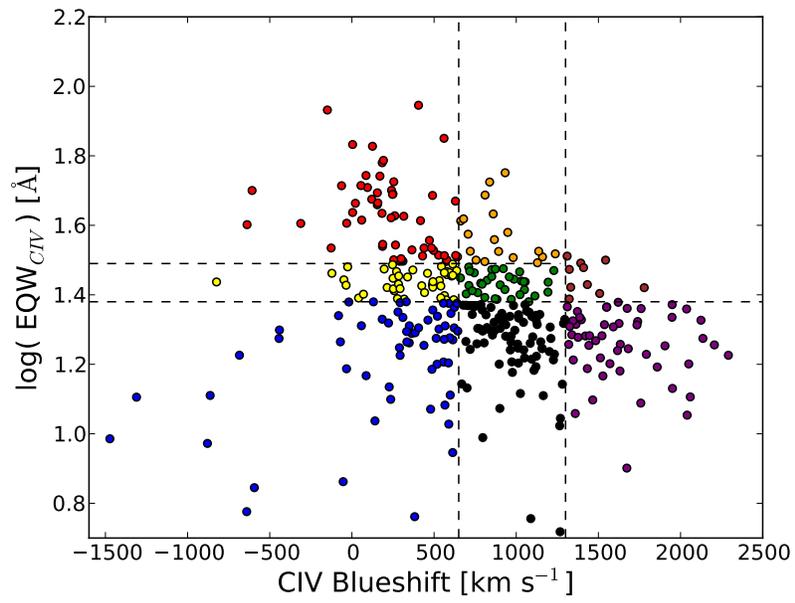}
\caption{Distribution of individual objects that make up the eight SEDs shown in Fig.~\ref{fig:sed}.  The color of the objects in each bin match the color of the SED they correspond to in Fig.~\ref{fig:sed}.}
\label{fig:sedpos}
\end{figure}

\begin{figure}[h!]
\epsscale{0.7}
\plotone{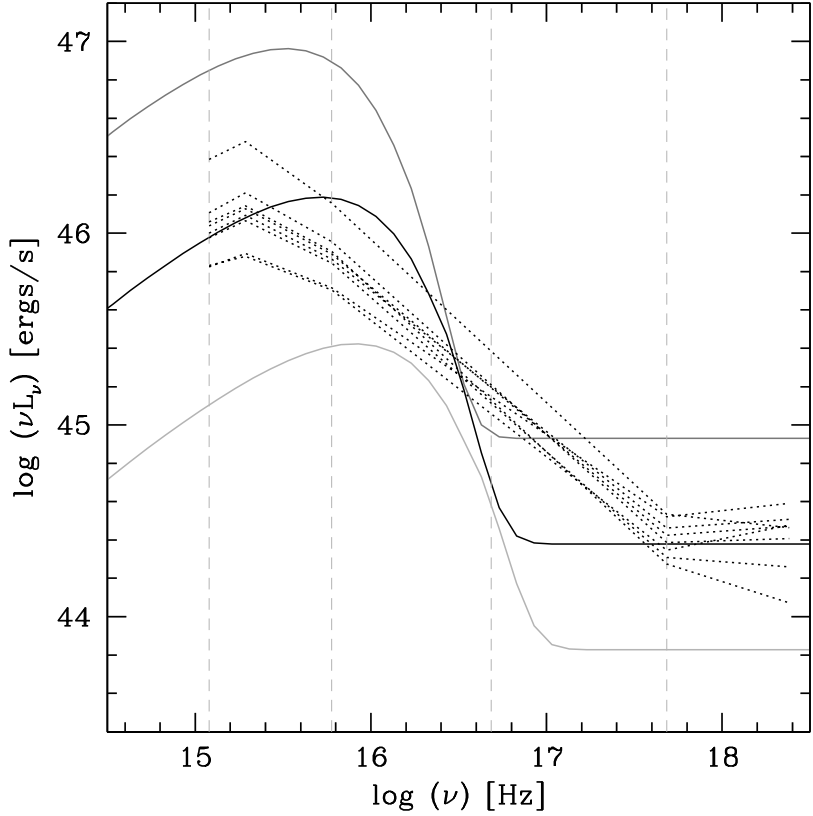}
\caption{Modified black-body SEDs.  The dark gray/black/light gray curves show SEDs with cutoff temperatures of 20/30/50\,eV.  Their luminosities are normalized in a way that is consistent with the $L_{\rm UV}$--$\alpha_{\rm ox}$ relationship from \citet{jbs+07}.  Note that the change in UV luminosity between the SEDs is greater than the change in X-ray luminosity.  The gray curves represent the extrema in terms of \aox, while the black curve represents the median.  We show the SEDs from Figure~\ref{fig:sed} (dotted curves) to highlight the differences in the far-UV spectral slope between the empirical and model SEDs and to emphasize the possible range of spectral shapes in the EUV.
As these are composite SEDs, their range is smaller than the extrema indicated by the gray model SEDs.  Vertical dashed lines indicate 2500\,\AA, 500\,\AA ($=24.8$\,eV), 0.2\,keV and 2\,keV.  
}
\label{fig:bb}
\end{figure}

Much as the FWHM of H$\beta$, strength of \ion{Fe}{2} (relative to H$\beta$), and soft X-ray properties can be used to divide low-$z$ quasars into extrema in the ``Eigenvector 1'' context \citep{bg92,lfe+97,sbm+07}, so can the properties of the \civ\ emission line at high redshift.  In \citet{Richards2011}, we showed that the extrema of large (small) \civ\ EQW and small (large) \civ\ blueshift can be attributed to different components of the BELR: a disk and a wind, respectively.  Here, we consider the UV through X-ray properties of radio-quiet quasars in the \civ\ EQW-blueshift parameter space and argue that these extrema are likely due to a hard ionizing spectrum in the ``disk''-dominated systems and a soft ionizing spectrum in the ``wind''-dominated systems (Figures~\ref{fig:aox},  \ref{fig:daox}, and \ref{fig:sed}).   Indeed these extrema could instead be considered as hard-spectrum quasars (HSQ) and soft-spectrum (SSQ) quasars.  More work is needed to understand the nature of quasars with weak \civ\ that appears at the systemic redshift and the relationship between the optical to X-ray flux ratio (\aox) and both the soft and hard X-ray spectral indices.  We argue that differences in the \civ\ properties of quasars could be used to better trace the structure of the SED in the two decades of frequency space between UV and X-ray measurements, allowing for more accurate bolometric corrections for individual quasars (Figures~\ref{fig:sed} and \ref{fig:bb}).
The \civ\ emission line may further allow a quantification the differences in the ``unseen'' part of the ionizing SED (e.g., an excess in the Helium continuum above the nominal assumed from \aox) and may help identify which objects are likely to show a soft X-ray excess feature.

\acknowledgments

Support for this work was provided by the National Aeronautics and Space Administration through Chandra Award Number G08-9103X issued by the Chandra X-ray Observatory Center, which is operated by the Smithsonian Astrophysical Observatory for and on behalf of the National Aeronautics Space Administration under contract NAS8-03060.  GTR acknowledges support from an Alfred P. Sloan Research Fellowship and NASA grant 07-ADP07-0035.  SCG thanks the National Science and Engineering Research Council of Canada and an Ontario Early Research Award for support.  KML acknowledges support by NSF AST-0707703.  Funding for the SDSS and SDSS-II has been provided by the Alfred P. Sloan Foundation, the Participating Institutions, the National Science Foundation, the U.S. Department of Energy, the National Aeronautics and Space Administration, the Japanese Monbukagakusho, the Max Planck Society, and the Higher Education Funding Council for England. The SDSS is managed by the Astrophysical Research Consortium for the Participating Institutions. The Participating Institutions are the American Museum of Natural History, Astrophysical Institute Potsdam, University of Basel, Cambridge University, Case Western Reserve University, University of Chicago, Drexel University, Fermilab, the Institute for Advanced Study, the Japan Participation Group, Johns Hopkins University, the Joint Institute for Nuclear Astrophysics, the Kavli Institute for Particle Astrophysics and Cosmology, the Korean Scientist Group, the Chinese Academy of Sciences (LAMOST), Los Alamos National Laboratory, the Max-Planck-Institute for Astronomy (MPIA), the Max-Planck-Institute for Astrophysics (MPA), New Mexico State University, Ohio State University, University of Pittsburgh, University of Portsmouth, Princeton University, the United States Naval Observatory, and the University of Washington.

\dataset [ADS/Sa.CXO#obs/9223] {Chandra ObsId 9223}
\dataset [ADS/Sa.CXO#obs/9224] {Chandra ObsId 9224}
\dataset [ADS/Sa.CXO#obs/9225] {Chandra ObsId 9225}
\dataset [ADS/Sa.CXO#obs/9226] {Chandra ObsId 9226}
\dataset [ADS/Sa.CXO#obs/9227] {Chandra ObsId 9227}
\dataset [ADS/Sa.CXO#obs/9228] {Chandra ObsId 9228}
\dataset [ADS/Sa.CXO#obs/9229] {Chandra ObsId 9229}
\dataset [ADS/Sa.CXO#obs/9230] {Chandra ObsId 9230}
\dataset [ADS/Sa.CXO#obs/9322] {Chandra ObsId 9322}
\dataset [ADS/Sa.CXO#obs/9323] {Chandra ObsId 9323}

\clearpage


\begin{thebibliography}{98}
\expandafter\ifx\csname natexlab\endcsname\relax\def\natexlab#1{#1}\fi
\expandafter\ifx\csname url\endcsname\relax
  \def\url#1{{\tt #1}}\fi
\expandafter\ifx\csname urlprefix\endcsname\relax\def\urlprefix{URL }\fi
\providecommand{\eprint}[2][]{\url{#2}}

\bibitem[\protect\astroncite{{Abazajian} et~al.}{2009}]{aaa+09}
{Abazajian}, K.~N., et~al. 2009, \apjs, 182, 543, \eprint{0812.0649}

\bibitem[\protect\astroncite{{Allen} et~al.}{2011}]{Allen10}
{Allen}, J.~T., {Hewett}, P.~C., {Maddox}, N., {Richards}, G.~T., \&
  {Belokurov}, V. 2011, \mnras, 410, 860, \eprint{1007.3991}

\bibitem[\protect\astroncite{{Arav} et al.}{1995}]{akb+95} Arav, N., Korista, K.~T., 
Barlow, T.~A., \& Begelman 1995, \nat, 376, 576 

\bibitem[\protect\astroncite{{Arnaud}}{1996}]{XSPEC}
{Arnaud}, K.~A. 1996, in Astronomical Data Analysis Software and Systems V, ed.
  {G.~H.~Jacoby \& J.~Barnes}, vol. 101 of {\em Astronomical Society of the
  Pacific Conference Series\/}, 17--+

\bibitem[\protect\astroncite{{Arnaud} et~al.}{1985}]{abc+85}
{Arnaud}, K.~A., et~al. 1985, \mnras, 217, 105

\bibitem[\protect\astroncite{{Avni} \& {Tananbaum}}{1982}]{at82}
{Avni}, Y. \& {Tananbaum}, H. 1982, \apjl, 262, L17

\bibitem[\protect\astroncite{{Bachev} et~al.}{2004}]{bms+04}
{Bachev}, R., {Marziani}, P., {Sulentic}, J.~W., {Zamanov}, R., {Calvani}, M.,
  \& {Dultzin-Hacyan}, D. 2004, \apj, 617, 171, \eprint{arXiv:astro-ph/0408334}

\bibitem[\protect\astroncite{{Baskin} \& {Laor}}{2004}]{bl04}
{Baskin}, A. \& {Laor}, A. 2004, \mnras, 350, L31,
  \eprint{arXiv:astro-ph/0403365}

\bibitem[\protect\astroncite{{Baskin} \& {Laor}}{2005}]{bl05}
--- 2005, \mnras, 356, 1029, \eprint{arXiv:astro-ph/0409196}

\bibitem[\protect\astroncite{{Blandford} \& {Payne}}{1982}]{bp82}
{Blandford}, R.~D. \& {Payne}, D.~G. 1982, \mnras, 199, 883

\bibitem[\protect\astroncite{{Boller} et~al.}{1996}]{bbf96}
{Boller}, T., {Brandt}, W.~N., \& {Fink}, H. 1996, \aap, 305, 53,
  \eprint{arXiv:astro-ph/9504093}

\bibitem[\protect\astroncite{{Boroson}}{2002}]{bor02}
{Boroson}, T.~A. 2002, \apj, 565, 78, \eprint{arXiv:astro-ph/0109317}

\bibitem[\protect\astroncite{{Boroson} \& {Green}}{1992}]{bg92}
{Boroson}, T.~A. \& {Green}, R.~F. 1992, \apjs, 80, 109

\bibitem[\protect\astroncite{{Brandt} et~al.}{1997}]{bme97}
{Brandt}, W.~N., {Mathur}, S., \& {Elvis}, M. 1997, \mnras, 285, L25,
  \eprint{arXiv:astro-ph/9703100}

\bibitem[\protect\astroncite{{Brotherton} \& {Francis}}{1999}]{bf99}
Brotherton, M.~S., \& Francis, P.~J.\ 1999, Quasars and Cosmology,
162, 395

\bibitem[\protect\astroncite{{Casebeer} et~al.}{2006}]{clb06}
{Casebeer}, D.~A., {Leighly}, K.~M., \& {Baron}, E. 2006, \apj, 637, 157,
  \eprint{arXiv:astro-ph/0508503}

\bibitem[\protect\astroncite{{Castor} et~al.}{1975}]{cak75}
{Castor}, J.~I., {Abbott}, D.~C., \& {Klein}, R.~I. 1975, \apj, 195, 157

\bibitem[\protect\astroncite{{Ciotti} et~al.}{2010}]{cop10}
{Ciotti}, L., {Ostriker}, J.~P., \& {Proga}, D. 2010, \apj, 717, 708,
  \eprint{1003.0578}

\bibitem[\protect\astroncite{{Collin} et~al.}{2006}]{Collin06}
{Collin}, S., {Kawaguchi}, T., {Peterson}, B.~M., \& {Vestergaard}, M. 2006,
  \aap, 456, 75, \eprint{arXiv:astro-ph/0603460}

\bibitem[\protect\astroncite{{Collin-Souffrin} et~al.}{1988}]{cdm+88}
{Collin-Souffrin}, S., {Dyson}, J.~E., {McDowell}, J.~C., \& {Perry}, J.~J.
  1988, \mnras, 232, 539

\bibitem[\protect\astroncite{{Crummy} et~al.}{2006}]{cfg+06}
{Crummy}, J., {Fabian}, A.~C., {Gallo}, L., \& {Ross}, R.~R. 2006, \mnras, 365,
  1067, \eprint{arXiv:astro-ph/0511457}

\bibitem[\protect\astroncite{{Done} \& {Nayakshin}}{2007}]{dc07}
{Done}, C. \& {Nayakshin}, S. 2007, \mnras, 377, L59,
  \eprint{arXiv:astro-ph/0701410}

\bibitem[\protect\astroncite{{Elvis} et~al.}{1994}]{ewm+94}
{Elvis}, M., et~al. 1994, \apjs, 95, 1

\bibitem[\protect\astroncite{{Everett}}{2005}]{everett05}
{Everett}, J.~E. 2005, \apj, 631, 689, \eprint{arXiv:astro-ph/0506321}

\bibitem[\protect\astroncite{{Ferland}}{2003}]{fer03}
{Ferland}, G.~J. 2003, \araa, 41, 517

\bibitem[\protect\astroncite{{Francis} et~al.}{1991}]{fhf+91}
{Francis}, P.~J., {Hewett}, P.~C., {Foltz}, C.~B., {Chaffee}, F.~H., {Weymann},
  R.~J., \& {Morris}, S.~L. 1991, \apj, 373, 465

\bibitem[\protect\astroncite{{Gallagher} et~al.}{2002}]{gbc+02}
{Gallagher}, S.~C., {Brandt}, W.~N., {Chartas}, G., \& {Garmire}, G.~P. 2002,
  \apj, 567, 37, \eprint{arXiv:astro-ph/0110579}

\bibitem[\protect\astroncite{{Gallagher} et~al.}{2006}]{gbc+06}
{Gallagher}, S.~C., {Brandt}, W.~N., {Chartas}, G., {Priddey}, R., {Garmire},
  G.~P., \& {Sambruna}, R.~M. 2006, \apj, 644, 709,
  \eprint{arXiv:astro-ph/0602550}

\bibitem[\protect\astroncite{{Gallagher} et~al.}{2005}]{grh+05}
{Gallagher}, S.~C., {Richards}, G.~T., {Hall}, P.~B., {Brandt}, W.~N.,
  {Schneider}, D.~P., \& {Vanden Berk}, D.~E. 2005, \aj, 129, 567,
  \eprint{arXiv:astro-ph/0410641}

\bibitem[\protect\astroncite{{Ganguly} et~al.}{2007}]{gbc+07}
{Ganguly}, R., {Brotherton}, M.~S., {Cales}, S., {Scoggins}, B., {Shang}, Z.,
  \& {Vestergaard}, M. 2007, \apj, 665, 990, \eprint{0705.1546}

\bibitem[\protect\astroncite{{Gaskell}}{1982}]{gas82}
{Gaskell}, C.~M. 1982, \apj, 263, 79

\bibitem[\protect\astroncite{{Gaskell}}{2009}]{gaskell09}
--- 2009, New Astronomy Review, 53, 140, \eprint{0908.0386}

\bibitem[\protect\astroncite{{Gibson} et~al.}{2008}]{gbs08}
{Gibson}, R.~R., {Brandt}, W.~N., \& {Schneider}, D.~P. 2008, \apj, 685, 773,
  \eprint{0808.2603}

\bibitem[\protect\astroncite{{Gierli{\'n}ski} \& {Done}}{2004}]{gd04}
{Gierli{\'n}ski}, M. \& {Done}, C. 2004, \mnras, 349, L7,
  \eprint{arXiv:astro-ph/0312271}

\bibitem[\protect\astroncite{{Goldschmidt} et~al.}{1999}]{gkm+99}
{Goldschmidt}, P., {Kukula}, M.~J., {Miller}, L., \& {Dunlop}, J.~S. 1999,
  \apj, 511, 612, \eprint{arXiv:astro-ph/9810392}

\bibitem[\protect\astroncite{{Green} \& {Mathur}}{1996}]{gm96}
{Green}, P.~J. \& {Mathur}, S. 1996, \apj, 462, 637,
  \eprint{arXiv:astro-ph/9512032}

\bibitem[\protect\astroncite{{Green}}{1996}]{Green96}
{Green}, P.~J. 1996, \apj, 467, 61, \eprint{arXiv:astro-ph/9603099}

\bibitem[\protect\astroncite{{Green} et~al.}{2004}]{gsc+04}
{Green}, P.~J., et~al. 2004, \apjs, 150, 43, \eprint{arXiv:astro-ph/0308506}

\bibitem[\protect\astroncite{{Green} et~al.}{2009}]{gar+09}
--- 2009, \apj, 690, 644, \eprint{0809.1058}

\bibitem[\protect\astroncite{{Hall} \& {Chajet}}{2010}]{hc10}
{Hall}, P.~B. \& {Chajet}, L.~S. 2010, in IAU Symposium, vol. 267 of {\em IAU
  Symposium\/}, 398--398

\bibitem[\protect\astroncite{{Hall} et~al.}{2006}]{hgr+06}
{Hall}, P.~B., {Gallagher}, S.~C., {Richards}, G.~T., {Alexander}, D.~M.,
  {Anderson}, S.~F., {Bauer}, F., {Brandt}, W.~N., \& {Schneider}, D.~P. 2006,
  \aj, 132, 1977, \eprint{arXiv:astro-ph/0606417}

\bibitem[\protect\astroncite{{Hewett} \& {Wild}}{2010}]{HW10}
{Hewett}, P.~C. \& {Wild}, V. 2010, \mnras, 405, 2302, \eprint{1003.3017}

\bibitem[\protect\astroncite{{Hopkins} et~al.}{2006}]{hhc+06}
{Hopkins}, P.~F., {Hernquist}, L., {Cox}, T.~J., {Di Matteo}, T., {Robertson},
  B., \& {Springel}, V. 2006, \apjs, 163, 1, \eprint{arXiv:astro-ph/0506398}

\bibitem[\protect\astroncite{{Isobe} et~al.}{1986}]{ifn86}
{Isobe}, T., {Feigelson}, E.~D., \& {Nelson}, P.~I. 1986, \apj, 306, 490

\bibitem[\protect\astroncite{{Just} et~al.}{2007}]{jbs+07}
{Just}, D.~W., {Brandt}, W.~N., {Shemmer}, O., {Steffen}, A.~T., {Schneider},
  D.~P., {Chartas}, G., \& {Garmire}, G.~P. 2007, \apj, 665, 1004,
  \eprint{0705.3059}

\bibitem[\protect\astroncite{{Kelly} et~al.}{2007}]{kbs+07}
{Kelly}, B.~C., {Bechtold}, J., {Siemiginowska}, A., {Aldcroft}, T., \&
  {Sobolewska}, M. 2007, \apj, 657, 116, \eprint{arXiv:astro-ph/0611120}

\bibitem[\protect\astroncite{{Konigl} \& {Kartje}}{1994}]{kk94}
{Konigl}, A. \& {Kartje}, J.~F. 1994, \apj, 434, 446

\bibitem[\protect\astroncite{{Korista} et~al.}{1997}]{kfb97}
{Korista}, K., {Ferland}, G., \& {Baldwin}, J. 1997, \apj, 487, 555,
  \eprint{arXiv:astro-ph/9704262}

\bibitem[\protect\astroncite{{Laor} et~al.}{1997}]{lfe+97}
{Laor}, A., {Fiore}, F., {Elvis}, M., {Wilkes}, B.~J., \& {McDowell}, J.~C.
  1997, \apj, 477, 93, \eprint{arXiv:astro-ph/9609164}

\bibitem[\protect\astroncite{{Leighly}}{1999}]{Leighly99}
{Leighly}, K.~M. 1999, \apjs, 125, 317, \eprint{arXiv:astro-ph/9907295}

\bibitem[\protect\astroncite{{Leighly}}{2001}]{lei01}
--- 2001, in Probing the Physics of Active Galactic Nuclei, ed.
  {B.~M.~Peterson, R.~W.~Pogge, \& R.~S.~Polidan}, vol. 224 of {\em
  Astronomical Society of the Pacific Conference Series\/}, 293,
  \eprint{arXiv:astro-ph/0012173}

\bibitem[\protect\astroncite{{Leighly}}{2004}]{Leighly04}
--- 2004, \apj, 611, 125, \eprint{arXiv:astro-ph/0402452}

\bibitem[\protect\astroncite{{Leighly} \& {Casebeer}}{2007}]{lc07}
{Leighly}, K.~M. \& {Casebeer}, D. 2007, in The Central Engine of Active
  Galactic Nuclei, ed. {L.~C.~Ho \& J.-W.~Wang}, vol. 373 of {\em Astronomical
  Society of the Pacific Conference Series\/}, 365--+

\bibitem[\protect\astroncite{{Leighly} et~al.}{2007}]{lhj+07}
{Leighly}, K.~M., {Halpern}, J.~P., {Jenkins}, E.~B., {Grupe}, D., {Choi}, J.,
  \& {Prescott}, K.~B. 2007, \apj, 663, 103, \eprint{arXiv:astro-ph/0611349}

\bibitem[\protect\astroncite{{Leighly} \& {Moore}}{2004}]{lm04}
{Leighly}, K.~M. \& {Moore}, J.~R. 2004, \apj, 611, 107,
  \eprint{arXiv:astro-ph/0402453}

\bibitem[\protect\astroncite{{Lucy} \& {Solomon}}{1970}]{ls70}
{Lucy}, L.~B. \& {Solomon}, P.~M. 1970, \apj, 159, 879

\bibitem[\protect\astroncite{{Marconi} et~al.}{2004}]{mrg+04}
{Marconi}, A., {Risaliti}, G., {Gilli}, R., {Hunt}, L.~K., {Maiolino}, R., \&
  {Salvati}, M. 2004, \mnras, 351, 169, \eprint{arXiv:astro-ph/0311619}

\bibitem[\protect\astroncite{{Marziani} et~al.}{1996}]{msd+96}
{Marziani}, P., {Sulentic}, J.~W., {Dultzin-Hacyan}, D., {Calvani}, M., \&
  {Moles}, M. 1996, \apjs, 104, 37

\bibitem[\protect\astroncite{{Marziani} et al.}{2010}]{msn+10} Marziani, P., 
Sulentic, J.~W., Negrete, C.~A., Dultzin, D., Zamfir, S., 
\& Bachev, R.\ 2010, \mnras, 409, 1033

\bibitem[\protect\astroncite{{Miller} et~al.}{2011}]{mbs+11}
{Miller}, B.~P., {Brandt}, W.~N., {Schneider}, D.~P., {Gibson}, R.~R.,
  {Steffen}, A.~T., \& {Wu}, J. 2011, \apj, 726, 20, \eprint{1010.4804}

\bibitem[\protect\astroncite{{Murray} et~al.}{1995}]{mcgv95}
{Murray}, N., {Chiang}, J., {Grossman}, S.~A., \& {Voit}, G.~M. 1995, \apj,
  451, 498

\bibitem[\protect\astroncite{{Osterbrock} \& {Pogge}}{1985}]{op85}
{Osterbrock}, D.~E. \& {Pogge}, R.~W. 1985, \apj, 297, 166

\bibitem[\protect\astroncite{{Porquet} et~al.}{2004}]{pro+04}
{Porquet}, D., {Reeves}, J.~N., {O'Brien}, P., \& {Brinkmann}, W. 2004, \aap,
  422, 85, \eprint{arXiv:astro-ph/0404385}

\bibitem[\protect\astroncite{{Pounds} et~al.}{1995}]{pdo95}
{Pounds}, K.~A., {Done}, C., \& {Osborne}, J.~P. 1995, \mnras, 277, L5

\bibitem[\protect\astroncite{{Press} et~al.}{1992}]{ptv+92}
{Press}, W.~H., {Teukolsky}, S.~A., {Vetterling}, W.~T., \& {Flannery}, B.~P.
  1992, {Numerical recipes in C. The art of scientific computing}

\bibitem[\protect\astroncite{{Proga}}{2003}]{Proga03}
{Proga}, D. 2003, \apj, 585, 406, \eprint{arXiv:astro-ph/0210642}

\bibitem[\protect\astroncite{{Proga} \& {Kallman}}{2004}]{pk04}
{Proga}, D. \& {Kallman}, T.~R. 2004, \apj, 616, 688,
  \eprint{arXiv:astro-ph/0408293}

\bibitem[\protect\astroncite{{Proga} et~al.}{1998}]{psd98}
{Proga}, D., {Stone}, J.~M., \& {Drew}, J.~E. 1998, \mnras, 295, 595

\bibitem[\protect\astroncite{{Proga} et~al.}{2000}]{psk00}
{Proga}, D., {Stone}, J.~M., \& {Kallman}, T.~R. 2000, \apj, 543, 686,
  \eprint{arXiv:astro-ph/0005315}

\bibitem[\protect\astroncite{{Reichard} et~al.}{2003}]{rrh+03}
{Reichard}, T.~A., et~al. 2003, \aj, 126, 2594, \eprint{arXiv:astro-ph/0308508}

\bibitem[\protect\astroncite{Richards et~al.}{2002}]{rvr+02}
Richards, G.~T., Berk, D. E.~V., Reichard, T.~A., Hall, P.~B., Schneider,
  D.~P., SubbaRao, M., Thakar, A.~R., \& York, D.~G. 2002, AJ, 124, 1

\bibitem[\protect\astroncite{Richards et~al.}{2003}]{rhv+03}
Richards, G.~T., et~al. 2003, AJ, 126, 1131

\bibitem[\protect\astroncite{Richards et~al.}{2006}]{rls+06}
--- 2006, ApJS, 166, 470

\bibitem[\protect\astroncite{{Richards} et~al.}{2011}]{Richards2011}
{Richards}, G.~T., et~al. 2011, \aj, 141, 167, \eprint{1011.2282}

\bibitem[\protect\astroncite{{Schmidt} \& {Green}}{1983}]{sg83}
{Schmidt}, M. \& {Green}, R.~F. 1983, \apj, 269, 352

\bibitem[\protect\astroncite{{Schneider} et~al.}{2010}]{srh+10}
{Schneider}, D.~P., et~al. 2010, \aj, 139, 2360, \eprint{1004.1167}

\bibitem[\protect\astroncite{{Scott} et~al.}{2004}]{Scott2004}
{Scott}, J.~E., {Kriss}, G.~A., {Brotherton}, M., {Green}, R.~F., {Hutchings},
  J., {Shull}, J.~M., \& {Zheng}, W. 2004, \apj, 615, 135,
  \eprint{arXiv:astro-ph/0407203}

\bibitem[\protect\astroncite {{Shen} et al.}{2008}]{sgs+08} Shen, Y., Greene, J.~E., 
Strauss, M.~A., Richards, G.~T., \& Schneider, D.~P.\ 2008, \apj, 680, 169 

\bibitem[\protect\astroncite{{Sobolewska} \& {Done}}{2007}]{sd07}
{Sobolewska}, M.~A. \& {Done}, C. 2007, \mnras, 374, 150,
  \eprint{arXiv:astro-ph/0609223}

\bibitem[\protect\astroncite{{Steffen} et~al.}{2006}]{ssb+06}
{Steffen}, A.~T., {Strateva}, I., {Brandt}, W.~N., {Alexander}, D.~M.,
  {Koekemoer}, A.~M., {Lehmer}, B.~D., {Schneider}, D.~P., \& {Vignali}, C.
  2006, \aj, 131, 2826, \eprint{arXiv:astro-ph/0602407}

\bibitem[\protect\astroncite{{Strateva} et~al.}{2005}]{sbs+05}
{Strateva}, I.~V., {Brandt}, W.~N., {Schneider}, D.~P., {Vanden Berk}, D.~G.,
  \& {Vignali}, C. 2005, \aj, 130, 387, \eprint{arXiv:astro-ph/0503009}

\bibitem[\protect\astroncite{{Sulentic} et~al.}{2000{\natexlab{a}}}]{smd00}
{Sulentic}, J.~W., {Marziani}, P., \& {Dultzin-Hacyan}, D. 2000{\natexlab{a}},
  \araa, 38, 521

\bibitem[\protect\astroncite{{Sulentic} et~al.}{2007}]{sbm+07}
{Sulentic}, J.~W., {Bachev}, R., {Marziani}, P., {Negrete}, C.~A., \&
  {Dultzin}, D. 2007, \apj, 666, 757, \eprint{0705.1895}

\bibitem[\protect\astroncite{{Sulentic} et~al.}{2000}]{szm+00}
{Sulentic}, J.~W., {Zwitter}, T., {Marziani}, P., \& {Dultzin-Hacyan}, D. 2000,
  \apjl, 536, L5, \eprint{arXiv:astro-ph/0005177}

\bibitem[\protect\astroncite{{Tytler} \& {Fan}}{1992}]{tf92}
{Tytler}, D. \& {Fan}, X. 1992, \apjs, 79, 1

\bibitem[\protect\astroncite{{Vanden Berk} et~al.}{2001}]{vrb+01}
{Vanden Berk}, D.~E., et~al. 2001, \aj, 122, 549,
  \eprint{arXiv:astro-ph/0105231}

\bibitem[\protect\astroncite{{Walter} \& {Fink}}{1993}]{wf93}
{Walter}, R. \& {Fink}, H.~H. 1993, \aap, 274, 105

\bibitem[\protect\astroncite{{Wang} et al.}{1996}]{wzg96} Wang, T.-G., Zhou, Y.-Y., 
\& Gao, A.-S.\ 1996, \apj, 457, 111 

\bibitem[\protect\astroncite{{Wang} et al.}{2011}]{wwz+11} Wang, H., Wang, T., Zhou, 
H., Liu, B., Wang, J., Yuan, W., \& Dong, X.\ 2011, arXiv:1106.2584 

\bibitem[\protect\astroncite{{Weymann} et~al.}{1991}]{wmf+91}
{Weymann}, R.~J., {Morris}, S.~L., {Foltz}, C.~B., \& {Hewett}, P.~C. 1991,
  \apj, 373, 23

\bibitem[\protect\astroncite{{Wilkes}}{1984}]{Wilkes84}
{Wilkes}, B.~J. 1984, \mnras, 207, 73

\bibitem[\protect\astroncite{{Wills} et~al.}{1999{\natexlab{a}}}]{wlb+99}
{Wills}, B.~J., {Laor}, A., {Brotherton}, M.~S., {Wills}, D., {Wilkes}, B.~J.,
  {Ferland}, G.~J., \& {Shang}, Z. 1999{\natexlab{a}}, \apjl, 515, L53

\bibitem[\protect\astroncite{{Wolf} et~al.}{2004}]{wmk+04}
{Wolf}, C., et~al. 2004, \aap, 421, 913, \eprint{arXiv:astro-ph/0403666}

\bibitem[\protect\astroncite{{Worrall} et~al.}{1987}]{wtg+87}
{Worrall}, D.~M., {Tananbaum}, H., {Giommi}, P., \& {Zamorani}, G. 1987, \apj,
  313, 596

\bibitem[\protect\astroncite{{Wu} et~al.}{2009}]{wvb+09}
{Wu}, J., {Vanden Berk}, D.~E., {Brandt}, W.~N., {Schneider}, D.~P., {Gibson},
  R.~R., \& {Wu}, J. 2009, \apj, 702, 767, \eprint{0907.2552}

\bibitem[\protect\astroncite{{Wu} et~al.}{2011}]{wbh+11}
{Wu}, J., et~al. 2011, ArXiv e-prints, \eprint{1104.3861}

\bibitem[\protect\astroncite{{York} et~al.}{2000}]{yaa+00}
{York}, D.~G., et~al. 2000, \aj, 120, 1579

\bibitem[\protect\astroncite{{Young} et~al.}{2009}]{yer09}
{Young}, M., {Elvis}, M., \& {Risaliti}, G. 2009, \apjs, 183, 17,
  \eprint{0905.0496}

\end{thebibliography}

\clearpage

\clearpage

\begin{deluxetable}{ccccccccc}
\rotate
\tablewidth{0pt}
\tablecaption{Archival Data\label{tab:tab1}}
\tablehead{
\colhead{Reference} &
\colhead{N Match} &
\colhead{N BAL} &
\colhead{N RL} &
\colhead{N $z$ cut} &
\colhead{N other cut} &
\colhead{N Kept\tablenotemark{a}} &
\colhead{$z_{\rm em}$} &
\colhead{$l_{\rm uv}$}
}
\startdata
Chandra C9 & 10 & 0 & 1 & 0 & 0 & 6 & 1.65--1.90 & 31.48--31.68 \\
Gallagher '05 & 14 & 1 & 2 & 0 & 3 & 10 & 1.69--2.17 & 31.00--31.73 \\
Strateva '05 & 155 &  0 & 10 & 94 & 89 & 23 & 1.55--2.24 & 30.35--31.54 \\
Kelly '07 & 157 & 1 & 9 & 95 & 112 & 6 & 1.57--2.03 & 30.07--30.93 \\
Just '07 & 32 & 3 & 10 & 26 & 21 & 5 & 1.91--2.27 & 32.05--32.23 \\
Gibson '08 & 536 & 52 & 71 & 100 & 210 & 181 & 1.70--2.31 & 30.34--32.11 \\
Green '09 & 281 & 11 & 29 & 146 & 182 & 46 & 1.54--2.26 & 30.49--31.79 \\
Young '09 & 792 & 33 & 78 & 423 & 604 & 91 & 1.55--2.27 & 30.10--31.75 \\
\enddata
\tablenotetext{a}{The number kept is determined by removing all of the BALs, low $z$ objects, and RL (or unmeasured radio) objects, but also by removing any duplicates between samples and objects failing to meet other cuts.
Thus the number of objects kept can be significantly smaller than the number of objects matched.}
\end{deluxetable}


\begin{deluxetable}{lcccccrl}
\tablewidth{0pt}
\tablecaption{New large \civ\ blueshift quasars observed with {\em Chandra}.\label{tab:chandra}}
\tablehead{
\colhead{Name} &
\colhead{$z_{\rm em}$} &
\colhead{$i$} &
\colhead{$\log L_{2800{\rm\AA}}$} &
\colhead{CIV shift} &
\colhead{CIV EQW} &
\colhead{Exptime} &
\colhead{OBSID} \\
\colhead{(SDSS J)} &
\colhead{} &
\colhead{} &
\colhead{(ergs/s/Hz)} &
\colhead{(km/s)} &
\colhead{(\AA)} &
\colhead{(ks)} &
\colhead{} 
}
\startdata
005102.42$-$010244.4 & 1.889 & 17.366 & 31.571 & 1597& 20.3 & 3.5 & 9224 \\
014812.23+000153.2 & 1.708 & 17.386 & 31.411 & 1590 & 15.2 & 10.5 & 9225 \\
020845.53+002236.0 & 1.898  & 16.723 & 31.854 & 1555 & 23.1 & 3.5 & 9223 \\
090007.14+321921.9 & 1.851 & 17.095 & 31.650 & 1955 & 21.6 & 11.0 & 9323 \\
100401.28+423123.0 & 1.666 & 16.733 & 31.725 & 2281 & 16.8 & 8.1 & 9322 \\
102907.06+651024.6 & 2.171 & 16.730 & 31.968 & 1560 & 17.2 & 10.6 & 9228 \\
115351.11+113649.2 & 1.681 & 17.262 & 31.463 & 2026 & 22.9 & 11.0 & 9230 \\
141949.39+060654.0 & 1.649 & 17.119 & 31.385 & 1941 & 23.5 & 9.9 & 9226 \\
150313.63+575151.6 & 1.721 & 17.075 & 31.593 & 1555 & 15.9 & 10.0 & 9227 \\
162622.06+295237.4 & 1.902 & 17.017 & 31.700 & 1501 & 20.6 & 10.9 & 9229
\enddata
\end{deluxetable}

\begin{deluxetable}{lcccccccr}
\tabletypesize{\footnotesize}
\rotate
\tablewidth{0pt}
\tablecaption{X-ray Properties
\label{tab:xcalc}
}
\tablehead{
\colhead{Name (SDSS J)} &
\colhead{\GHR\tablenotemark{a}} &
\colhead{$\log(F_{\rm X})$\tablenotemark{b}} &
\colhead{$\log(f_{\rm 2 keV})$\tablenotemark{c}} &
\colhead{$\log(f_{\rm 2500})$\tablenotemark{c}} &
\colhead{$\log(L_{\rm 2500})$\tablenotemark{d}} &
\colhead{\aox} &
\colhead{\daox} 
%
}
\startdata
005102.42$-$010244.4 &  2.75$^{+0.85}_{-0.62}$ & $-13.354\pm0.085$ & $-30.754\pm0.085$ & $-26.425$ & 31.510 & $-1.66$ & $-0.01$ \\
014812.23+000153.2    &  2.13$^{+0.32}_{-0.28}$ & $-13.340\pm0.052$ & $-30.980\pm0.052$ & $-26.426$ & 31.434 & $-1.75$ & $-0.09$ \\
020845.53+002236.0    &  1.62$^{+0.28}_{-0.26}$ & $-12.818\pm0.055$ & $-30.678\pm0.055$ & $-26.170$ & 31.768 & $-1.73$ & $-0.03$ \\
090007.14+321921.9    &  2.20$^{+0.25}_{-0.25}$ & $-13.229\pm0.043$ & $-30.818\pm0.043$ & $-26.286$ & 31.635 & $-1.74$ & $-0.05$ \\ 
100401.28+423123.0    &  1.65$^{+0.27}_{-0.27}$ & $-13.181\pm0.056$ & $-31.050\pm0.056$ & $-26.158$ & 31.677 & $-1.88$ & $-0.19$ \\
102907.06+651024.6    &  2.10$^{+0.23}_{-0.20}$ & $-13.083\pm0.038$ & $-30.662\pm0.038$ & $-26.074$ & 31.971 & $-1.76$ & $-0.03$ \\
115351.11+113649.2    &  2.10$^{+0.23}_{-0.20}$ & $-13.111\pm0.039$ & $-30.770\pm0.039$ & $-26.361$ & 31.484 & $-1.69$ & $-0.02$ \\
141949.39+060654.0    &  2.65$^{+0.35}_{-0.28}$ & $-13.252\pm0.044$ & $-30.742\pm0.044$ & $-26.295$ & 31.533 & $-1.71$ & $-0.04$ \\
150313.63+575151.6    &  2.30$^{+0.23}_{-0.23}$ & $-13.101\pm0.038$ & $-30.676\pm0.038$ & $-26.292$ & 31.574 & $-1.68$ & $0.00$ \\
162622.06+295237.4    &  2.60$^{+0.40}_{-0.30}$ & $-13.373\pm0.048$ & $-30.811\pm0.048$ & $-26.264$ & 31.679 & $-1.75$ & $-0.06$
\enddata
\tablenotetext{a}{\GHR\ is a coarse measure of the hardness of the X-ray spectrum determined by comparing the observed \HR\ to a simulated \HR\ that takes into account spatial and temporal variations in the instrument response.}
\tablenotetext{b}{The full-band X-ray flux, $F_{\rm X}$, has units of \flux\ and is calculated by integrating the power-law spectrum given by $\Gamma$ and normalized by the full-band count rate from 0.5--8.0~keV.  The errors are derived from the 1$\sigma$ errors in the full-band count rate.} 
\tablenotetext{c}{X-ray and optical flux densities were measured at rest-frame 2~keV and 2500\,\AA, respectively; units are \fnu.}
\tablenotetext{d}{The 2500\,\AA\ luminosity density, $L_{\rm 2500}$, has units of \lumin~Hz$^{-1}$.}
\end{deluxetable}

\end{document}